\documentclass[aps,pra,superscriptaddress,showpacs,tightenlines,cap,cs5size,winfonts,nospace,indent,fancyhdr,twocolumn]{revtex4-1}
\usepackage{amsmath}
\usepackage{changepage}
\usepackage{epsfig}
\usepackage{epstopdf}
\usepackage[colorlinks,citecolor=blue,linkcolor=blue]{hyperref}
\usepackage{graphicx}
\usepackage{lipsum}

\begin{document}

\title{Simultaneous blockade of a photon phonon, and magnon induced by a
two-level atom }
\author{Chengsong Zhao}
\affiliation{School of Physics, Dalian University of Technology, Dalian
    116024,China}
\author{Xun Li}
\affiliation{School of Physics, Dalian University of Technology, Dalian
    116024,China}
\affiliation{National Key Laboratory of Shock Wave and
Detonation Physics, Institute of Fluid Physics, China Academy of Engineering
Physics, Mianyang 621900, China}
\author{Shilei Chao}

\affiliation{School of Physics, Dalian University of Technology, Dalian
    116024,China}
\author{Rui Peng}
\affiliation{School of Physics, Dalian University of Technology, Dalian
    116024,China}
\author{Chong Li}
\affiliation{School of Physics, Dalian University of Technology, Dalian
    116024,China}

\author{Ling Zhou}
\email{zhlhxn@dlut.edu.cn}
\affiliation{School of Physics, Dalian University of Technology, Dalian
    116024,China}

\begin{abstract}
The hybrid microwave optomechanical-magnetic system has recently emerged as
a promising candidate for coherent information processing because of the
ultrastrong microwave photon-magnon coupling and the longlife of the magnon
and phonon. As a quantum information processing device, the realization of
single excitation holds special meaning for the hybrid system. In this paper,
we introduce a single two-level atom into the optomechanical-magnetic
system and show that an unconventional blockade due to destructive interference cannot offer a blockade of both the photon and magnon. Meanwhile under the condition of single excitation resonance,
the blockade of photon, phonon, and magnon can be achieved simultaneously
even in a weak optomechanical region, but the phonon blockade still requires the cryogenic temperature condition.
\end{abstract}

\maketitle

\section{Introduction}

The effect of one photon preventing the second photon entrance is called a
photon blockade \cite{Birnbaum,PhysRevLett.79.1467}, which is the pivotal
effect to achieve photons at the quantum level. It is believed that
photon blockade can be used as a single photon source and to process quantum
information \cite{hacker2016photon}. The photon blockade in the cavity-QED systems %
\cite{PhysRevA.87.023822,PhysRevA.91.043831,PhysRevA.95.063842} were thoroughly
investigated and have been achieved in experiments \cite%
{Dayan1062,PhysRevLett.121.043601}. Recently, the optomechanical system has
attracted significant attention, such as working as a sensor to detect tiny
mass and force \cite%
{PhysRevLett.97.133601, PhysRevLett.105.123601,PhysRevLett.113.020405,
Zhang_2017} , a platform to investigate the fundamental physics \cite
{PhysRevLett.116.163602} and a device to processing quantum information \cite
{PhysRevLett.109.013603,Wang_2012,PhysRevLett.111.083601,
dongchunhua,PhysRevLett.109.193602,PhysRevA.91.063836,LiXun}. The most
attractive characteristic of an optomechanical system is the nonlinearity
resulting from the radiation pressure, which can induce Kerr nonlinearity 
\cite{PhysRevLett.107.063601} and produce the photon blockade \cite
{PhysRevA.88.023853}. However, currently, the single-photon optomechanical
coupling is still within a weak coupling region, which only induces only fainter Kerr nonlinearity.
Therefore, some strategies were put forward to enhance the nonlinearity \cite
{PhysRevA.88.063854,PhysRevLett.109.063601}. To avoid the weakness of the
delicate single-photon nonlinear coupling, the photon blockade resulting from
destructive interference called unconventional blockade (UB) was proposed
and thoroughly investigated \cite{PhysRevA.87.013839,PhysRevLett.121.043602}.

Most recently, the photon-magnon coupling system in the microwave \cite
{PhysRevLett.113.156401, PhysRevB.93.144420, PhysRevApplied.2.054002} and
optical frequency \cite{PhysRevLett.117.123605,PhysRevLett.117.133602,
PhysRevLett.116.223601} regime has aroused attention. Different from
the weak optomechanical coupling, the ultrastrong coupling between microwave
photons and magnons [the collective spin excitation in yttrium iron garnet
(YIG)] was realized\cite{PhysRevB.93.144420,niemczyk2010circuit}, and the
magnons possess a very low damping rate. Meanwhile, the magnon excitation
interacting with phonons (vibrational modes of the YIG sphere) is similar to
the optomechanical interaction, \cite{PhysRevLett.121.203601}, so, both kinds of interactions magnetic-mechanical \cite{zhang2016cavity} and optical-mechanical are nonlinear. The phonons and magnons posse coupling
mediated by cavity fields \cite{PhysRevA.96.023826}, and the entanglement of a
magnon, photon, and phonon in cavity magnomechanics has been investigated where
photon-magnon and magnomechanical interactions were considered \cite
{PhysRevLett.121.203601}. In Ref. \cite{PhysRevA.100.043831}, the supermode of a 
photon exhibits blockade under the Kerr effect in optomagnonic microcavities
system.

The photon blockade can be generated from the destructive interference \cite
{PhysRevA.96.053810, PhysRevA.87.013839} as well as the single excitation
resonance \cite{PhysRevA.90.023849, PhysRevLett.107.063601,
PhysRevB.87.235319, PhysRevA.99.043837, PhysRevA.95.063842}. Usually, the
destructive interference and the single excitation resonance resulting from
dressed states can supply a better blockade than the Kerr effect because of
the weak coupling strength of the Kerr interaction. The photon blockade in an
optomechanical system \cite{PhysRevA.87.013839} as well as in an optomagnonic
system \cite{PhysRevA.100.043831} were separately thoroughly investigated. A magnon
blockade via qubit-magnon coupling has been studied in Ref. \cite
{PhysRevB.100.134421}. However, in the hybrid optomechanical-magnetic
system, the simultaneous blockade of the photon, phonon, and magnon has not
been studied. Meanwhile, the hybrid system has special significance for the realization of quantum information processing, like the quantum internet \cite{PhysRevA.100.023801}. If the hybrid optomechanical-magnetic system was used as a
quantum device, the single excitation level is important, and the
simultaneous blockade of photon, phonon, and magnon should be pivotal and deserves
further investigation.

In this paper, we consider a hybrid microwave optomechanical-magnetic system aiming to generate the simultaneous photon-phonon-magnon blockade.
Considering the achievement of ultrastrong microwave optical-magnetic
coupling in experiments \cite{PhysRevLett.113.156401, PhysRevLett.111.127003}%
, we derive three-partite interaction among photon, phonon, and magnon. By
introducing a single two-level atom, under the condition of single
excitation resonance, we show that the simultaneous blockade of
photon, phonon, and magnon can be achieved with the assistance of the
three-partite interaction on the condition of cryogenic temperature of the mechanical mode, while the unconventional destructive interference
can not offer the simultaneously multi-modes antibunching. In our scheme the
single-photon strong optomechanical coupling is not required, therefore, it
can be feasible in experiment. Our scheme is a guideline for hybrid
optomechanical-magnetic experiments nearing the regime of single-photon
nonlinearity, and for potential quantum information processing applications
with photons, magnons, and phonons.

\section{The model and the analytical analysis}

\begin{figure}[tbp]
\centering  \includegraphics[width=0.44\textwidth]{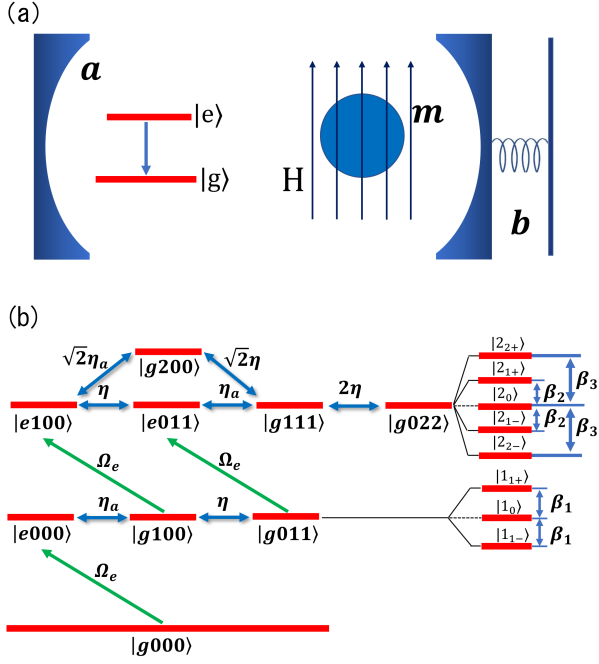}
\caption{(a) Sketch of the system. A two-level atom is placed inside a
microwave cavity with a movable mirror. A YIG sphere is placed near the
maximum magnetic field of the cavity mode, and in a uniform bias magnetic field, which establishes the
magnon-photon coupling. (b) Energy-level diagram under the Hamiltonian Eq.
\eqref{heff}, where $|g(e),n_{+},n_{-},n_{b}\rangle $ denotes lower-level $g$
(upper level $e$), and $n_{j}(j=+,-,b)$ is the number of the mode $(a_{+}, a_{-}, b)$
. The eigenstates are drawn on the right-hand side.}
\label{fig1}
\end{figure}

We consider a hybrid optomechanical-magnetic system, where a two-level atom
and a YIG microsphere are contained in the microwave cavity, and one of the
mirrors is movable, shown in Fig.~\ref{fig1}(a). The magnons are sourced
from a collective spins in a ferrimagnet. Here, we ignore the interaction
between magnons and phonons due to deformation of the YIG sphere, because
the single-magnon magnomechanical coupling rate is typically small \cite
{PhysRevLett.121.203601, PhysRevA.96.023826}. The magnetic dipole mediates
the coupling between magnons and cavity photons. The Hamiltonian of the
system reads

\begin{equation}
H=H_{om}+H_{op}+H_{ao}+H_{d},  \label{eq:h}
\end{equation}
where
\begin{eqnarray}
H_{om} &=&\omega _{c}a^{\dagger }a+\omega _{m}m^{\dagger }m+G_{m}(a^{\dagger
}m+am^{\dagger }),  \notag \\
H_{op} &=&\omega _{b}b^{\dagger }b+ga^{\dagger }a(b^{\dagger }+b),  \notag \\
H_{ao} &=&\omega _{a}\sigma ^{\dagger }\sigma +g_{a}(\sigma a^{\dagger
}+\sigma ^{\dagger }a), \\
H_{d} &=&\Omega _{e}(\sigma e^{i\omega _{L}t}+\sigma ^{\dagger }e^{-i\omega
_{L}t}),  \notag
\end{eqnarray}
$j^{\dagger }$($j,j=a,m,b$) is the creation (annihilation) operator of the
related mode (photon, magnon, and phonon) with frequency $\omega _{c}$, $
\omega _{m}$ and $\omega _{b}$, respectively. $\sigma $ stands for the 
pseudo-spin of the two-level atom. $H_{om}$ consists of the energy of the
photon and magnon, as well as the photon-magnon interaction with the effective
strength $G_{m}$, which is called the cavity magnon polaritons \cite%
{PhysRevLett.120.057202}. $H_{op}$ is composed of the energy of the phonon and
the optomechanical interaction with coupling strength $g$. The first term in
$H_{ao}$ is the energy of the atom, and the second term describes the atom
interacting with the cavity field. $H_{d}$ denotes an atom pumped with a 
classical field with frequency $\omega _{L}$.

\begin{figure}[tbp]
\centering\includegraphics[width=0.48\textwidth]{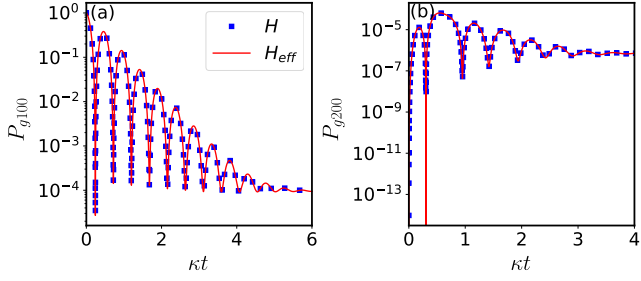}
\caption{The evolution of probabilities $P_{{g100}}$ (a) and $P_{{g200}}$
(b) with original Hamiltonian $H$ (red line) and effective Hamiltonian $%
H_{eff}$ (blue squares), respectively, where $P_{{g100}}=|C_{g100}|^{2}$, $%
P_{{\ g200}}=|C_{g200}|^{2}$. The parameters are $\protect\eta=5\protect%
\kappa $, $\protect\eta_{a}=6/\protect\sqrt{2} \protect\kappa $, $G_{m}=200%
\protect\kappa $, and $\Omega _{e}=0.1 \protect\kappa $.}
\label{fig2}
\end{figure}

In the frame rotating with $H_{0}=\omega _{L}(a^{\dagger }a+\sigma ^{\dagger
}\sigma +m^{\dagger }m)$, the Hamiltonian can be changed into
time-independent. For simplicity, we assume $\omega _{m}=\omega _{c}$, then $
\delta =\omega _{c(m)}-\omega _{L}$. We diagonalize the Hamiltonian $
H_{0}^{^{\prime }}=\delta (a^{\dagger }a+m^{\dagger }m)+G_{m}(a^{\dagger
}m+am^{\dagger })$ by introducing supermodes $a_{\pm }=\frac{1}{\sqrt{2}}%
(a\pm m)$. Considering photon-magnon interaction larger than the
optomechanical and atom-photon interaction, i.e., $G_{m}$ $\gg \{g,g_{a}\}$
and choosing $\omega _{b}=2G_{m}$, we rewrite the Hamiltonian as
\begin{eqnarray}
H_{eff} &=&\Delta a_{+}^{\dagger }a_{+}+(\Delta -2G_{m})a_{-}^{\dagger
}a_{-}+\omega _{b}b^{\dagger }b+\Delta _{a}\sigma ^{\dagger }\sigma  \notag
\\
&&-\eta (a_{+}^{\dagger }a_{-}b+a_{+}a_{-}^{\dagger }b^{\dagger })+\eta
_{a}(a_{+}^{\dagger }\sigma +a_{+}\sigma ^{\dagger })  \notag \\
&&+\Omega _{e}(\sigma +\sigma ^{\dagger }),  \label{heff}
\end{eqnarray}%
where $\Delta =\delta +G_{m}$, $\eta =g/2$, $\eta _{a}=g_{a}/\sqrt{2}$, $
\Delta _{a}=\omega _{a}-\omega _{L}$. The detailed deduction of Hamiltonian~%
\eqref{heff} is given in Appendix~\ref{appendix}. For simplicity, hereafter
we will assume $\Delta =\Delta _{a}$. We see that the effective Hamiltonian
contains three-partite interaction, which is similar to in Ref. \cite
{PhysRevA.87.013839}. Differently from their scheme, we introduce a pumped
two-level atom aiming to achieve a blockade of the photon, magnon, and phonon. We
also would like to compare the different effect of a blockade between the
destructive interference mechanism and the single excitation resonance
mechanism. Observe the last two brackets in Eq.~\eqref{heff}; the pumped
two-level atom interacts with mode $a_{+}$, which results in the blockade of
mode $a_+$. Although the three-partite nonlinear interaction means the
parametric-down conversion form between $a_{-}$ and $b$ mediated by
absorption or emission of mode $a_{+}$, the blockade of the mode $a_{+}$ can
not result in the amplification in mode $a_{-}$ and $b$. Instead, if there
is only one excitation in the mode $a_{+}$, the transfer of the single
excitation creates only one excitation in every mode of $a_{-}$ and $b$,
that is to say, the blockade in mode $a_{+}$ will lead to the blockade in
mode $a_{-}$ and mode $b$; therefore it is possible to generate a blockade in
supermodes $a_{+}$, $a_{-}$ and mode $b$. We will show that the bare modes $%
a $, $b$, and $m$ can also be blockaded simultaneously.\newline
\indent To check the validity of the approximation from Hamiltonian~%
\eqref{eq:h} to Hamiltonian~\eqref{heff}, we choose $|g100\rangle $ as the
 initial state and plot the evolution of the probabilities of states $%
|g200\rangle $ and $|g100\rangle $ governed by the Hamiltonians $H$ and $%
H_{eff}\ $respectively, shown in Fig.~\ref{fig2}, where $%
|e(g),n_{+},n_{-},n_{b} \rangle $ represents a state with atom in $|e\rangle
$ ($|g\rangle $), and $|n_{+}\rangle$, $|n_{\_}\rangle$, and $|n_{b}\rangle $
are the number state for the $a_{+}$, $a_{-}$, and $b$ modes, respectively.
From Fig.~\ref{fig2}, we see clearly that the results of original
Hamiltonian agree very well with that of effective Hamiltonian $H_{eff}$,
which means that the effective Hamiltonian $H_{eff}$ is reliable.\newline
\indent Due to the limit of the weak driving field, for understanding the
blockade mechanism of the photon (phonon, magnon), we temporarily ignore the
pumping of the atom and derive the eigenstates and eigenvalues of $H_{eff}$~%
\eqref{heff} in the few-photon subspace, yielding
\begin{equation}
\begin{split}
|0\rangle :& \lambda _{0}=0, \\
|1_{0}\rangle :& \lambda _{10}=\Delta , \\
|1_{\pm }\rangle :& \lambda _{1\pm }=\Delta \pm \beta _{1}, \\
|2_{0}\rangle :& \lambda _{20}=2\Delta , \\
|2_{1\pm }\rangle :& \lambda _{21\pm }=2\Delta \pm \beta _{2}, \\
|2_{2\pm }\rangle :& \lambda _{22\pm }=2\Delta \pm \beta _{3},
\end{split}
\label{eq:resonance}
\end{equation}
where $\beta _{1}=\sqrt{\eta _{a}^{2}+\eta ^{2}}$, $\beta _{2}=\sqrt{\frac{
3\eta _{a}^{2}+7\eta ^{2}-D}{2}}$, $\beta _{3}=\sqrt{\frac{3\eta
_{a}^{2}+7\eta ^{2}+D}{2}}$, $D=\sqrt{\eta _{a}^{4}+26\eta _{a}^{2}\eta
^{2}+25\eta ^{4}}$. The expression of the dressed states $|s_{c}\rangle $ ($%
s=0,1,2$; $c=0$,$\pm ,1\pm ,2\pm )$ is given in Appendix \ref{appendix}, and
the energy-levels are shown on the right side of Fig.~\ref{fig1}(b).

In the weak driving limit, to analytically derive the equal-time second-order
correction function, the state of the system can be truncated in few
excitation subspace and approximately expressed as
\begin{equation}
\begin{split}
|\psi \rangle =& C_{g000}|g000\rangle +C_{g100}|g100\rangle
+C_{g011}|g011\rangle \\
& +C_{e000}|e000\rangle +C_{g200}|g200\rangle +C_{g111}|g111\rangle \\
& +C_{e100}|e100\rangle +C_{g022}|g022\rangle +C_{e011}|e011\rangle .
\end{split}
\label{eq:psi}
\end{equation}%
Under the action of the non-Hermite Hamiltonian $\widetilde{H}
=H_{eff}-i(\kappa _{+}a_{+}^{\dagger }a_{+}+\kappa _{-}a_{-}^{\dagger
}a_{-}+\kappa _{a}\sigma ^{\dagger }\sigma )$ with the decay rate $\kappa
_{j}$ ($j=+,-,a$), the probability amplitude in $|\psi \rangle $ can be
obtained by solving the Schr\"{o}dinger equation $i\partial |\psi \rangle
/\partial t=\widetilde{H}|\psi \rangle $. The detail of the deduction and
the steady-state solution can be found in Appendix~\ref{solution_equation}.

To characterize nonclassical photon (magnon, phonon) statistics, we employ and 
equal-time second-order correlation function defined by
\begin{equation}
g_{i}^{2}(0)=\frac{\rm Tr(c_{i}^{\dagger }c_{i}^{\dagger }c_{i}c_{i}\rho )}{\rm 
[Tr(c_{i}^{\dagger }c_{i}\rho )]^{2}},  \label{eq:correlation original}
\end{equation}%
where $i=a,m,b,a_{+},a_{-}$. The steady-state correlation functions of our
system can be analytically obtained via the steady-state wave function 
\eqref{eq:psi} as
\begin{eqnarray}
g_{a_{+}}^{2}(0) &=&\frac{2|C_{g200}|^{2}}{(|C_{g100}|^{2}+u_1)^{2}}\approx
\frac{2|C_{g200}|^{2}}{|C_{g100}|^{4}}\ ,  \label{ga+} \\
g_{a_{-}}^{2}(0) &=&\frac{2|C_{g022}|^{2}}{
(|C_{g011}|^{2}+u_2)^2}\approx \frac{2|C_{g022}|^{2}}{
|C_{g011}|^{4}},  \notag  \label{eq:g20}
\end{eqnarray}%
with $u_1=2|C_{g200}|^{2}+|C_{g111}|^{2}+|C_{e100}|^{2}$, and $u_2=2|C_{g022}|^2+|C_{g111}|^2+|C_{e011}|^2$ where the second
approximate equals in Eq.~\eqref{ga+} are obtained under the conditions $
|C_{g000}|\gg \{|C_{g100}|$, $|C_{g011}|$, $|C_{e000}|\}\gg$ $\{|C_{g200}|$,
$|C_{g111}|$, $|C_{g022}|$, $|C_{e100}|$, $|C_{e011}|\}$. For mode $b$ , it
is not reasonable to obtain $g_{b}^{2}(0)$ with the analytical solution \eqref{eq:psi} because its decay has been ignored. We will directly
calculate it from the master equation. The correlation function $g_{i}^{2}(0)\geq 1$ is referred to as Poissonian and super-Poissonian. The correlation function $g_{i}^{2}(0)<1 $
indicates sub-Poissonian, and the limit $g_{i}^{2}(0) \rightarrow 0$
corresponds to the complete blockade. Remarkably, the single-photon regime
is usually characterized by $g_{i}^{2}(0)<0.5$ \cite{PhysRevA.96.053810}. From the
expression Eq.~\eqref{ga+} and Eq.~\eqref{steadsolution}, one can see that
the blockade in mode $a_{+}$ ($a_{-}$) is possible only if the
population $C_{g200}$ $(C_{g022})\approx 0$. We will plot second-order
correlation function and discuss it further in the next section.

Although the polariton modes \cite{PhysRevLett.120.057202,PhysRevB.98.174423, zhang2017observation} consisting
of optical mode and magnetic mode can be indirectly derived by directly
detecting the output spectrum of photons, the blockade of the photon, phonon, and
magnon still deserve our investigation. Due to the combination of the optical
mode and magnetic mode, the statistical properties of supermodes $a_{\pm }$
and bare modes $a$ and $m$ are different. In order to see clearly the
difference, we derive the relations between the two bases
\begin{equation}
\begin{split}
& |00\rangle _{d}=|00\rangle , \\
& |10\rangle _{d}=\frac{1}{\sqrt{2}}(|10\rangle +|01\rangle ), \\
& |01\rangle _{d}=\frac{1}{\sqrt{2}}(|10\rangle -|01\rangle ), \\
& |02\rangle _{d}=\frac{1}{2}(|20\rangle -\sqrt{2}|11\rangle +|02\rangle ),
\\
& |20\rangle _{d}=\frac{1}{2}(|20\rangle +\sqrt{2}|11\rangle +|02\rangle ),
\\
& |11\rangle _{d}=\frac{1}{\sqrt{2}}(|20\rangle -|02\rangle ),
\end{split}
\label{eq:relationship}
\end{equation}%
where the left side states are labeled by $|n_{+},n_{-}\rangle _{d}$ ($n_{+}$
and $n_{-}\ $correspond to the Fock state of mode $a_{+}$ and $a_{-}$) while
right-side state are labeled with$|n_{m},n_{a}\rangle $ ($n_{m}$ and $n_{a}\
$denote the Fock state of mode $m$ and $a$). The derivation of Eq.~\eqref{eq:relationship} is given in Appendix~\ref{appendix_3}. See the last
line in Eq.~\eqref{eq:relationship}, where the state $|11\rangle _{d}$ means only
one excitation in mode $a_{+}$ and $a_{-}$, however for the modes $m$ and $a$
, they might be populated in two excitations. That is to say, the blockade
of supermodes $a_{+}$ and $a_{-}$ does not mean the blockade of bare modes $%
a $ and $m$. Therefore, we need to calculate the second-order correlation of
the mode $m$ and $a$:
\begin{equation}
\begin{split}
& g_{a}^{2}(0)\approx \frac{2(|C_{g200}|^{2}+|C_{g022}|^{2}+2|C_{g111}|^{2})%
}{(|C_{g100}|^{2}+|C_{g011}|^{2})^{2}}, \\
& g_{m}^{2}(0)=g_{a}^{2}(0).
\end{split}
\label{eq:g20a}
\end{equation}
We can see that the correlation functions for the optical and magnetic mode are
the same. The blockades in the modes $m$ and $a$ require that $C_{g200}$, $%
C_{g022}$ and $C_{g111}$ reach zero simultaneously. Fortunately, as one can
observe from Eq. ~\eqref{steadsolution}, when $C_{g022}$ equals zero, $%
C_{g111}$ is equal to zero too. That is to say, when both $a_{+}$ and $a_{-}$
modes are a blockade, the photon and magnon modes $a$ and $m$ are both a 
blockade too.

\section{The statistical properties of the multimode field}

In the above analytical calculation of $g^2_{i}(0)$ $(i=a_{\pm},b,a,m)$, we have
made some approximations. We now show the correction of the approximations
and investigate the statistical properties of the multimode field. For simplicity, we assume that the decay rates of the 
optical mode, magnetic mode, and atom are equal, and then we can derive the
master equation as
\begin{equation}
\begin{split}
\dot{\rho}=& -i[H_{eff},\rho ]+\kappa (\mathcal{D}[a_{+}]+\mathcal{D}%
[a_{-}]+ \mathcal{D}[\sigma ])\rho \\
& +(n_{th}+1)\kappa _{b}\mathcal{D}[b]\rho + n_{th} \kappa _{b}\mathcal{D}[b^{\dagger}]\rho,
\end{split}
\label{eq:master}
\end{equation}
where $\rho$ is the density matrix of the hybrid system, $\mathcal{D}%
[o]\rho =2o\rho o^{\dagger }-o^{\dagger }o\rho -\rho o^{\dagger}o$, and $n_{th}$ is the thermal phonon population. We assume
that the average particle numbers of photons (magnons) in thermal equilibrium are zero because of their high frequencies.
\begin{figure}[h]
\centering 
\includegraphics[width=0.48\textwidth]{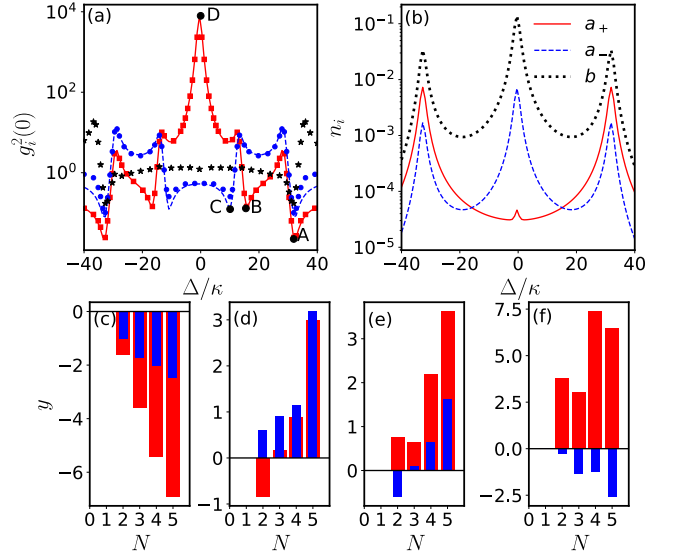}
\caption{(a) Equal-time second-order correlation function for modes $a_{+}$ (red solid, red square), $
a_{-}$ (blue dashed, blue dot), and $b$ (black asterisk) versus detuning $\Delta $, where lines
and marks represent analytical and numerical solutions respectively.
(b) The average particle number. The relative probability population
function $y$ (c)-(f) corresponding to the mark point A-D in (a), where red (light gray) and blue (dark gray) bars represent supermode $a_+$  and $a_-$, respectively. The parameters are $\protect%
\eta _{a}=40/\protect\sqrt{2}\protect\kappa $, $\protect\eta =15\protect%
\kappa $, $\protect\kappa _{b}=0.05\protect\kappa $, $\Omega =0.1\protect%
\kappa $, $G_{m}=800\protect\kappa $, and $n_{th}=0$.}
\label{fig3}
\end{figure}

We truncate the Fock space up to $|5\rangle $ for modes $a_{\pm }$ and $b$.
Based on the subspace consisting of the two-level atom and the modes $a_{\pm
} $ and $b$, we numerically solve Eq.~\eqref{eq:master} and calculate the
second-order correlation function of mode $a_{\pm }$ and $b$. In Fig.~\ref%
{fig3}(a), we plot $g_{a_{\pm }}^{2}(0)$ with an analytical solution of Eqs.~%
\eqref{ga+} and numerical results of Eq.~\eqref{eq:master}, respectively. We
see that they agree well, which means that we can understand the second
order correlation with the analytic analysis. In order to make clear the
relation between the mechanism of blockade and the probability distribution,
we define the function $y(N)={\rm log}_{10}^{{}}\frac{P(N)}{P_{p}(N)}$ where $P(N)$
is the probability in $|N\rangle $, and $P_{p}(N)$ is Poissonian distribution;
thus the value of $y$ reveals the relative difference between the population
and Poissonian distribution. In Fig.~\ref{fig3}(c) to \ref{fig3}(f), we plot $y(N)$
corresponding to point A to D respectively. If $y$ is positive, population
at $N$ excitation is higher than Poissonian distribution, or otherwise it is
lower than the Poissonian distribution.

For the mark point A in Fig.~\ref{fig3}(a), $\Delta =\beta _{1}$, $\lambda
_{1-}=0$, which means the single excitation resonance. Then $|1_{-}\rangle $ can be easily populated (for the symmetry
point of A, $\Delta =-\beta _{1}$, $\lambda _{1+}=0$, then $|1_{+}\rangle $
is easy to be populated ). Notice the expression $|1_{-}\rangle $ in Eq.~%
\eqref{dress}, where there is only one excitation in modes $a_{\pm }$ and $b$, so
we can see strong blockade in $a_+$, $a_-$, and $b$ modes under the same
condition. Meanwhile the average numbers $n_{a_{+}}$, $n_{a_{-}}$, and $n_{b}$
reach their local maximum of $n_{a_{\pm }(b)}$[see Fig.~\ref{fig3}(b)]. All
of the probability at $N>1$ is less than Poissonian distribution due to the 
resonance mechanism, shown in Fig.~\ref{fig3}(c).

For the mark point B in Fig.~\ref{fig3}(a), $g_{a_{+}}^{2}(0)$ achieves a 
local minimum value where the real part of numerator of $C_{g200}\ $is zero.
By observing Fig.~\ref{fig1}(b), the two jumps $|e100\rangle \rightarrow
|g200\rangle $ and $|g111\rangle \rightarrow |g200\rangle $ destructively
interfere each other, such that the population in $|g200\rangle $ is low, so the
mode $a_{+}$ is blockade. That is the so-called UB. However, under this
condition, the $a_{\_}$ mode is super-Poissonian because there is a population
in $|g111\rangle $, resulting in population $|g022\rangle $. By observing
Fig.~\ref{fig3} (d), the destructive interference only decreases the
probability in $N=2$ for the mode $a_{+}$, while for the mode $a_{-}$ the
probability for $N>1$ is higher than Poissonian distribution. This result
indicates that the destructive interference can not offer blockade for both supermodes $a_{+}$ and $a_{-}$.

For the point C in Fig.~\ref{fig3}(a), $g_{a_{\_}}^{2}(0)$ achieves a local
minimum value. As one can observe from Eq.~\eqref{steadsolution}, the
requirement for $C_{g022}\approx 0$ is the same as that for $C_{g111}\approx
0$, if \{$\eta ,\eta _{a}\}\neq 0$. As seen in Fig.~\ref{fig1}(b), there are
two jumps $|g200\rangle \rightarrow |g111\rangle $ and $|e011\rangle
\rightarrow |g111\rangle $. Their destructive interference results in
blockade in mode $a_{-}$. Meanwhile, there is a population in the state $
|g200\rangle $, which means the super-Poissonian in mode $a_{+}$.
Correspondingly, in Fig.~\ref{fig3}(e), the population of mode $a_{+}$ is
still higher than the Poissonian distribution, while for the mode $a_{-}$, the
destructive interference only decrease the probability in only $N=2$. This result
is similar to what we have pointed in the analysis of point B, i.e., the
destructive interference can not offer us a simultaneous blockade in supermode
$a_{+}$ and $a_{-}.$

For the point D, in Fig.~\ref{fig3}(a), $\Delta =0$, $\lambda _{10}=\lambda
_{20}=0$, which means that the single excitation resonance $|1_{0}\rangle $
and double resonant excitation $|2_{0}\rangle $ are both satisfied.
Observing Eq. ~\eqref{dress}, the resonance between state $|1_{0}\rangle $
and state $|0\rangle $ can lead to the populations in the states $%
|g011\rangle $ and $|e000\rangle $. Likewise, the population in $%
|2_{0}\rangle $ means that the states $|g200\rangle $, and $|e011\rangle $
are easily populated too, while the state $|g022\rangle $ is not so easily
populated because of the mutual cancellation between $\eta $ and $\eta _{a}$
[the factor $\frac{\eta _{a}^{2}-\eta ^{2}}{\sqrt{2}\eta ^{2}A_{1}}$ is
smaller than $\frac{\eta _{a}}{\beta _{1}}$, see Eq.~\eqref{dress}].
Therefore, the mode $a_{+}$ will be strong super-Poissonian, and the mode $%
a_{-}$ is sub-Poissonian. The results are corresponding to Fig.~\ref{fig3}%
(f), where the population for mode $a_{+}$ is higher than the Poissonian
distribution, and the probabilities distribution for mode $a_{-}$ are less
than Poissonian.

\begin{figure}[t]
    \centering\includegraphics[width=0.48\textwidth]{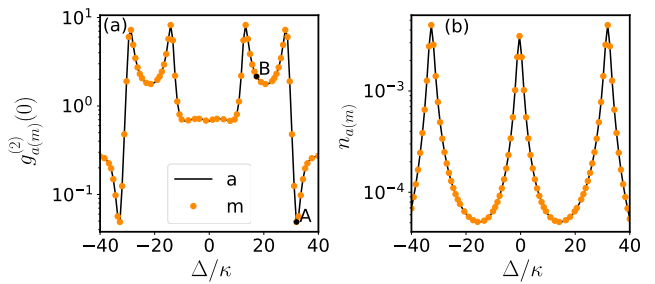}
    \caption{(a): Equal time second correlation function for modes $a$ (solid line), $%
    m $ (dots). (b): average number for optical mode $a$ (solid line) and magnetic
    mode $m$ (dots) as functions of detuning $\Delta $. The other parameters are
    the same as in Fig.~\ref{fig3}.}
    \label{fig4}
\end{figure}

As we have mentioned before, for mode $b$, $g_{b}^{2}(0)$ should not be
calculated from an analytical solution Eq.~\eqref{eq:psi}. We directly
calculate $g_{b}^{2}(0)$ with the master equation \eqref{eq:master}, shown in
Fig.~\ref{fig3}(a). We see that around point A, we can also achieve blockade
in mode $b$. Therefore, under single excitation resonance, all of the modes $%
a_{+}$, $a_{\_}$, and $b$ exhibit the blockade phenomenon. In addition, the
parameters, in Fig.~\ref{fig3}, $\frac{g^{2}}{\omega _{b}\kappa } =9/16 < 1$
means that weak photon nonlinearity from radiation pressure in an optomechanical system could generate a photon, magnon, and phonon blockade, in our system. But the single-photon optomechanical
coupling $g$ is still larger than the damping rate $\kappa $. We will show
that the single excitation resonant does not require $g>\kappa $; that is to
say, even under the condition $g<\kappa $, we still can obtain the
simultaneous blockade for the three modes.

In Fig.~\ref{fig4}, we plot $g_{a}^{2}(0)$ (solid) and $g_{m}^{2}(0)$ (dots), where, 
obviously, they are the same and agree well with Eq.~\eqref{eq:g20a}. As we
have analyzed before,  the blockade of $a_{+}$ mode means $|C_{g200}|^{2}$ $%
\approx 0$, and the $a_{-}$ mode blockade corresponds to $|C_{g022}|^{2}\approx 0$
(also $|C_{g111}|^{2}\approx 0$). When both $a_{+}$ and $a_{-}$ modes are
a blockade, from the expression Eq ~\eqref{eq:g20a}, the photon $a$ and the
magnon $m$ are both blockade; therefore at point A [see Fig.~\ref{fig4}(a)],
the optical mode and magnetic mode are both a blockade. However, around $%
\Delta =0$ (point D), the statistical property of $g_{a_{+}}^{2}(0)$ is
different from that of $g_{a_{-}}^{2}(0)$, $g_{a(m)}^{2}(0)$ still showing
sub-Poissonian. From Eqs.~\eqref{eq:g20}, \eqref{eq:g20a}, and 
\eqref{steadsolution}, we obtain
\begin{equation}
g_{a}^{2}(0)\approx \frac{1}{F_{2}^{2}}g_{a_{+}}^{2}(0)+(\frac{2}{F_{1}}-%
\frac{1}{F_{1}^{2}})g_{a_{-}}^{2}(0),  \label{eq:ga_ap}
\end{equation}%
where $F_{1}=|\frac{\tilde{\Delta}}{\eta }|^{2}+1$, $F_{2}=|\frac{\eta }{%
\tilde{\Delta}}|^{2}+1$, and $\tilde{\Delta}=\Delta -i\kappa $. So, when $%
\Delta $ is extremely small, $F_{1}\rightarrow 1$ and $F_{2}\rightarrow
\infty $, then, $g_{a}^{2}(0)$ is dominated by $g_{a_{-}}^{2}(0)$.
Therefore, we can observe a sub-Poissonian around $\Delta =0$ regime.
Comparing the value of $g_{a(m)}^{2}(0)$ around point B with that around
point A, we see that the sub-Poissonian resulting from destructive
interference (point B) does not exist, but the blockade resulting from single
excitation resonance (point A) still exists.

\begin{figure}
    \centering \includegraphics[width=0.48\textwidth]{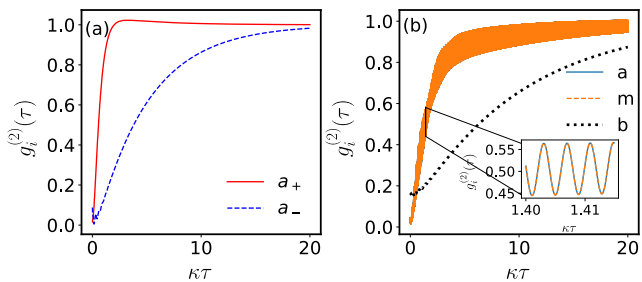}
    \caption{Time-delay second-order correlation function for supermode $a_{+}$ (red solid line) and $a_{-}$ (blue dashed line) in (a), and for optical mode (blue solid line), magnetic mode
    (orange dashed line), and mechanical mode (black dotted line) in (b). We set $\Delta =%
    \protect\beta _{1}$, and other parameters are the same as in Fig.~\ref{fig3}. The
    panel in (b) shows the partial enlarged detail.}
    \label{fig5}
\end{figure}

To further characterize the blockade of modes $a_{\pm}$, $b$, $a$, and $m$, 
choosing a single excitation resonance condition $\Delta =\beta _{1}$, we plot a 
second-order delay correlation function defined by $g_{i}^{(2)}(\tau )=\frac{%
\langle c_{i}^{\dagger }(0)c_{i}^{\dagger }(\tau )c_{i}(\tau
)c_{i}(0)\rangle }{\langle c_{i}^{\dagger }(0)c_{i}(0)\rangle ^{2}}$ in Fig.~
\ref{fig5}. $g_{i}^{2}(\tau )\leq g_{i}^{2}(0)$ is called bunching, and $%
g_{i}^{2}(\tau )>g_{i}^{2}(0)$ is called antibunching which is also the
quantum signature \cite{Scully:1601132}. Meanwhile, $g^{2}(\tau )$ is
proportional to the condition probability for detecting a second photon
(magnon, phonon) at $t=\tau $, given that a photon (magnon, phonon) has been
detected earlier at $t=0$ \cite{PhysRevA.90.063824}. Observing Figs.~\ref
{fig5}(a) and (b), because of the single excitation resonance, the
time-delay correction functions for supermodes $a_{\pm }$ and optical,
magnetic, or mechanical mode are all antibunching even in the weak photon nonlinear
region. $g_{m}^{(2)}(\tau )$ agrees well with $g_{a}^{(2)}(\tau )$ which is
just like the equal-time second-order correlation function. Comparing Figs.~\ref
{fig5}(a) and (b), the time-delay correction function of supermodes $a_{\pm} $ has no quick oscillations, but that of the optical and magnetic mode
exhibits quick oscillations. The quick local oscillations in the time-delay
second-order function for optical and magnetic mode results from the
interference between supermodes $a_{+}$ and $a_{-}$ and the frequency of mechanical mode $b$ \cite{PhysRevA.90.023849}.

\begin{figure}[tbp]
    \centering \includegraphics[width=0.48\textwidth]{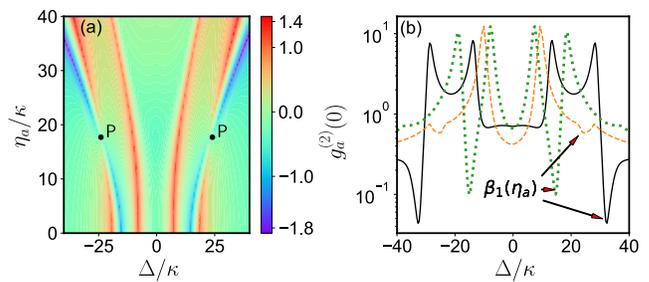}
    \caption{ (a): Contour plot $log_{10}^{{}}g_{a}^{2}(0)$ as function of $%
    \protect\eta _{a}$ and $\Delta $. (b): $g_{a}^{2}(0)$ change with $\Delta $
    for several values of $\protect\eta _{a}=40/\protect\sqrt{2}\protect\kappa $ 
    (black solid line), $17.7\protect\kappa $ (orange dashed line), $0.5\protect\kappa $ 
    (green dotted line). The other parameters are the same as in Fig.~\ref{fig3}}
    \label{fig6}
\end{figure}

We now investigate the second-order correlation function $g_{a(m)}^{2}$
affected by the coupling strength $\eta_a$ shown in Fig.~\ref{fig6}. From
Figs.~\ref{fig6}(a) and (b), we observe that with the increasing of $\eta
_{a} $, the low value $log_{10} ^{{}} g^{(2)}_a(0)$ points (single
excitation resonance) in terms of $\Delta $ are increased, which is because
the resonant condition $\Delta =\beta _{1}$ is increased with $\eta _{a}$.
In Fig.~\ref{fig6}(b), interestingly, the minimum value of $g_{a(m)}^{2}$ is
not monotonous decreasing with increasing $\eta _{a}$. When $\eta
_{a}\approx 17.7\kappa $, $g_{a(m)}^{2}$ is abnormal where the effect of the
single excitation resonance does not result in a blockade as in the other case. See
the mark point P in Fig.~\ref{fig6}(a), where there is a cross where the $\Delta
=\beta _{1}$ (the single excitation resonance ) and $\Delta =\beta _{2}/2$
(two excitation resonance) are both satisfied, so, $g_{a(m)}^{2}$ can not
show a blockade. Except for the cross point, the larger value of $\eta _{a}$, the
better the blockade.

\begin{figure}[htp]
    \centering\includegraphics[width=0.48\textwidth]{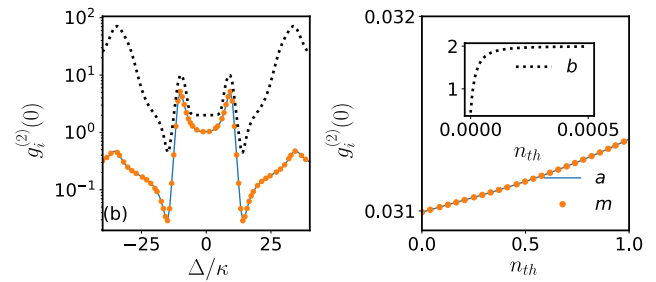}
    \caption{(a) Equal-time second-order correlation function for mode $a$ (blue solid line), $m$ (orange dots), and $b$ (black dotted line). We set $\protect
    \eta =0.2\kappa$, $\Omega _{e}=0.8\protect\kappa $, $\protect\eta _{a}=20/\protect
    \sqrt{2}\kappa$, and $\protect\kappa _{b}=\protect\kappa $ in (a). The other parameters
    are same as in Fig.~\ref{fig3}. (b) Equal-time second-order correlation function versus thermal phonon population. We set $\Delta=\beta_1$, and other parameters are the same as in panel (a).}
    \label{fig_add}
\end{figure}

We now show that it is possible to generate a photon, magnon, and phonon
blockade without a strong optomechanical coupling coefficient. In Fig.~\ref%
{fig_add}, both $\frac{g^{2}}{\omega _{b}\kappa }\ll 1$ and $g<\kappa $ are satisfied, and we plot the equal-time second-order correlation function for
modes $b$, $a$, and $m$. In Fig.~\ref{fig_add}(a), due to single
excitation resonance, the strong sub-Poissonian for modes $a$, $m$, and $b$
can be observed, while the destructive interference resulting in a blockade is
not observed in the weak coupling regime. Here, although the single-photon
optomechanical coupling is small, the large atom-photon interaction $g_{a}$
makes $\beta _{1}$ larger than $\kappa $, which ensures the blockade of the photon, magnon, and phonon. We can understand it from Eq.(\ref{steadsolution}). In order to keep single excitation, the denominator of $
C_{g100}$ ($C_{g011,}C_{e000}$) should be as low as possible, i.e., min%
$|\widetilde{\Delta }^2 -\eta ^{2}-\eta _{a}^{2}|$, then we
deduce the condition $\Delta =\sqrt{\eta ^{2}+\eta _{a}^{2}-\kappa ^{2}}$ .
Therefore, even $\eta <\kappa $, the relative large value of $\eta _{a}$
still can make $\eta ^{2}+\eta _{a}^{2}>\kappa ^{2}$, and then the single
excitation will dominate the wave function, and the blockade can be obtained.
We can conclude that the single excitation resonance can result in a multimode blockade even in a weak optomechanical coupling region while the destructive interference can not offer us multimode antibunching.

In Fig.~\ref{fig_add}(b), we plot the equal-time second-order correlation functions of a  photon, magnon, and phonon affected by thermal phonon number. As we can observe the blockade of a photon and magnon under a weak coupling regime still exists after considering the thermal environment of a phonon, but the phonon blockade disappears and the  correlation function approaches 2 with increasing $n_{th}$. When the thermal phonon population is taken into account, the state of the system truncated the in few excitation subspace can be expressed as mixed a state of $|\psi_n\rangle$ \cite{PhysRevA.87.013839} where
\begin{equation}
    \begin{split}
    |\psi_n \rangle =& C_{g00n}|g00n\rangle +C_{g10n}|g10n\rangle
    +C_{g0n+1}|g01n+1\rangle \\
    & +C_{e00n}|e00n\rangle +C_{g20n}|g20n\rangle +C_{g11n+1}|g11n+1\rangle \\
    & +C_{e10n}|e10n\rangle +C_{g02n+2}|g02n+2\rangle\\
     &+C_{e01n+1}|e01n+1\rangle.
    \end{split}
    \label{eq:psin}
\end{equation}
Because of the three-partite interaction $a_{+} a_-^{\dagger} b^{\dagger} +h.c.$, the thermal phonon cannot be converted into a photon and magnon. From  Eq.~\eqref{eq:psin}, although the thermal phonon can be in the state $|n\rangle$, the states of photon and magnon still can be in $|0\rangle$ or $|1\rangle$, which means the blockade of modes $a$ and $m$ still exists, but phonon blockade will be destroyed $(n>1)$ \cite{PhysRevA.93.063861, PhysRevB.100.134421, PhysRevA.100.053802,PhysRevA.94.063853}, and the correlation function of the phonon will close to the that of thermal field. But, the blockade of the photon and magnon is affected slightly by the thermal environment because of the change in the single excitation resonance for $|\psi_n\rangle$ \cite{PhysRevA.87.013839}. Therefore, to generate simultaneous blockade of a photon, phonon, and magnon, the small thermal phonon population is necessary.

\section{Discussion and conclusion}
When the single excitation resonance condition is satisfied,
The simultaneous blockade of a photon, phonon and magnon can offer us some potential applications. The usual hybrid system mainly contains two different physical systems, but the quantum internet may require more complex quantum information processing, like the processing and storing of information while simultaneously updating the information in a quantum information circuit and network \cite{PhysRevA.100.023801}. The simultaneous blockade of a multimode field could be used in this process and be more powerful than the  usual single mode blockade. If we realize the single excitation, from Eq.~\eqref{eq:relationship}, the photon and magnon will be a in Bell state $1/\sqrt{2}(|10\rangle \pm |01\rangle)$, which is useful in quantum information processing.

\indent From Fig.~\ref{fig3} to Fig.~\ref{fig_add}, the parameter $G_{m}$
is seemingly not important in numerical simulation, but we do need strong
magnon-photon coupling, because we require the condition $G_{m}\gg \{\eta
,\eta _{a}\}$ to achieve the effective Hamiltonian, and the three-partite
interaction is true only under this condition. Recently, strong and even
ultrastrong coupling between photons and magnons at microwave frequencies,
using of a YIG sphere, has been reported \cite
{PhysRevLett.113.156401,PhysRevB.93.144420}. For instance, in Ref. \cite
{PhysRevLett.113.156401}, the magnon-photon coupling strength was achieved
as high as ${\rm g=2\pi \times 2.5GHz}$, and dissipation rates of the microwave
photon and the magnon resonance are ${\rm \kappa _{a}=2\pi \times 33MHz}$ and ${\rm \kappa _{m}=2\pi \times 15MHz}
$, respectively. Currently, the optomechanical
single-photon strong-coupling condition $g>\kappa$ is still a challenge.
Most of the experiments of the optomechanical system are still within the 
single-photon weak coupling regime \cite{Nature2011,dongchunhua,Clark}. In
our scheme, the three-partite interaction is results from the
optomechanical interaction, but the single-photon strong coupling is not necessary.

In this paper, we put forward a scheme to generate a photon, phonon and magnon
blockade in a hybrid microwave optomechanical-magnetic system. By introducing
a two-level atom interacting with the cavity field, we carefully compare the
blockade resulting from destructive interference and that resulting from
single excitation resonance. We find that the blockade resulting from single
excitation resonance is much better than that resulting from destructive
interference. Most importantly, under the same detuning condition, the
photon, phonon and magnon can be blockade simultaneously. Furthermore, we find that the phonon blockade is easy to be destroyed by thermal excitation, while the blockade of the  photon and magnon are affected slightly by the thermal environment. To generate simultaneous blockade of the photon, phonon and magnon, the small thermal phonon population is necessary.    \newline
\indent In our system, the multipartite interaction results from
optomechanical coupling, which is the key factor to obtain the simultaneous
blockade of the photon, phonon, magnon. However, the single excitation is the
condition of the simultaneous blockade, and the single-photon strong
optomechanical coupling condition is not required. Therefore, the present
scheme is feasible in experiment, which is a guideline for hybrid
optomechanical-magnetic experiments nearing the regime of single-photon
nonlinearity, and for potential quantum information processing applications
with photons, magnons and phonons.

\section*{Acknowledgements}

We are grateful to J.Q. You and Guo-Qiang Zhang for enlightening discussions.
This work was supported by NSFC under Grant No. 11874099. \appendix

\section{The derivation of an effective Hamiltonian and its eigenstates}

\label{appendix} In this appendix, we give the detailed derivation for
Hamiltonian~\eqref{heff}. In the frame rotating with $H_{0}=\omega
_{L}(a^{\dagger }a+\sigma ^{\dagger }\sigma +m^{\dagger }m)$, the
Hamiltonian~ \eqref{eq:h} can be written as
\begin{eqnarray}
H &=&\delta _{c}a^{\dagger }a+\delta _{m}m^{\dagger }m+G_{m}(a^{\dagger
}m+am^{\dagger })  \notag \\
&&+\omega _{b}b^{\dagger }b+ga^{\dagger }a(b^{\dagger }+b)  \notag \\
&&+\Delta _{a}\sigma ^{\dagger }\sigma +g_{a}(\sigma a^{\dagger }+\sigma
^{\dagger }a) \\
&&+\Omega _{e}(\sigma +\sigma ^{\dagger }),  \notag
\end{eqnarray}%
with $\delta _{c(m)}=\omega _{c(m)}-\omega _{L}$. For simplicity, we assume $
\omega _{m}=\omega _{c}$, then $\delta _{c}=\delta _{m}=\delta $. We
diagonalize the Hamiltonian $H_{0}^{^{\prime }}=\delta (a^{\dagger
}a+m^{\dagger }m)+G_{m}(a^{\dagger }m+am^{\dagger })$ by introducing $a_{\pm
}=\frac{1}{\sqrt{2}}(a\pm m)$, then $H_{0}^{^{\prime }}=(\delta
+G_{m})a_{+}^{\dagger }a_{+}+(\delta -G_{m})a_{-}^{\dagger }a_{-}$. Choosing
$H_{f0}=\Delta a_{+}^{\dagger }a_{+}+(\Delta -2G_{m})a_{-}^{\dagger
}a_{-}+\omega _{b}b^{\dagger }b+\Delta _{a}\sigma ^{\dagger }\sigma $ and
assuming $\omega _{b}=2G_{m}$, $\Delta =\Delta _{a}$, we switch into the
interaction picture and obtain
\begin{equation}
\begin{split}
H_{I}=\eta & (a_{+}^{\dagger }a_{+}+a_{-}^{\dagger }a_{-}-a_{+}^{\dagger
}a_{-}e^{i2G_{m}t}-a_{-}^{\dagger }a_{+}e^{-i2G_{m}t}) \\
\times & (be^{-i\omega _{b}t}+b^{\dagger }e^{i\omega _{b}t})+\Omega
_{e}(\sigma e^{-i\Delta t}+\sigma ^{\dagger }e^{i\Delta t}) \\
& +\eta _{a}(a_{+}^{\dagger }\sigma +a_{+}\sigma ^{\dagger }-a_{-}^{\dagger
}\sigma e^{-i2G_{m}t}-a_{-}\sigma ^{\dagger }e^{i2G_{m}t}),
\end{split}
\end{equation}
where $\eta _{a}=\frac{g_{a}}{\sqrt{2}}$, $\eta =\frac{g}{2}$. The detuning $
\Delta $ can be arbitrary value. Considering $G_{m}\gg \{\eta ,\eta _{a}\}$,
we take rotating wave approximation and ignore high frequency terms, then the 
Hamiltonian could be written as
\begin{equation}
\begin{split}
H_{I}=& -\eta (a_{+}^{\dagger }a_{-}b+a_{-}^{\dagger }a_{+}b^{\dagger
})+\eta _{a}(a_{+}^{\dagger }\sigma +a_{+}\sigma ^{\dagger }) \\
& +\Omega _{e}(\sigma e^{i\Delta t}+\sigma ^{\dagger }e^{-i\Delta t}).
\end{split}%
\end{equation}%
We would like to rewrite the Hamiltonian into time-independent form by
switching back into original picture, then we have
\begin{equation}
\begin{split}
H_{eff}=& \Delta a_{+}^{\dagger }a_{+}+(\Delta -2G_{m})a_{-}^{\dagger
}a_{-}+\omega _{b}b^{\dagger }b+\Delta _{a}\sigma ^{\dagger }\sigma \\
& -g_{a}/2(a_{+}^{\dagger }a_{-}b+a_{-}^{\dagger }a_{+}b^{\dagger }) \\
& +g/\sqrt{2}(a_{+}^{\dagger }\sigma +a_{+}\sigma ^{\dagger })+\Omega
_{e}(\sigma +\sigma ^{\dagger }).
\end{split}%
\end{equation}%
It is exactly the effective Hamiltonian~\eqref{heff}.

In the limit of a weak driving field, we temporary forget the pumping of the
atom and derive the eigenstates and eigenvalues of $H_{eff}$ in the
few-photon subspace, yielding
\begin{equation}
\begin{split}
|0\rangle :& \lambda _{0}=0, \\
|1_{0}\rangle :& \lambda _{10}=\Delta , \\
|1_{\pm }\rangle :& \lambda _{1\pm }=\Delta \pm \beta _{1}, \\
|2_{0}\rangle :& \lambda _{20}=2\Delta , \\
|2_{1\pm }\rangle :& \lambda _{21\pm }=2\Delta \pm \beta _{2}, \\
|2_{2\pm }\rangle :& \lambda _{22\pm }=2\Delta \pm \beta _{3},
\end{split}%
\end{equation}%
where $\beta _{1}=\sqrt{\eta _{a}^{2}+\eta ^{2}}$, $\beta _{2}=\sqrt{\frac{
3\eta _{a}^{2}+7\eta ^{2}-D}{2}}$, $\beta _{3}=\sqrt{\frac{3\eta
_{a}^{2}+7\eta ^{2}+D}{2}}$, $D=\sqrt{\eta _{a}^{4}+26\eta _{a}^{2}\eta
^{2}+25\eta ^{4}}.$ The corresponding eigenstates are
\begin{equation}
\begin{split}
|0\rangle =& |g000\rangle, \\
|1_{0}\rangle =& \frac{1}{\beta _{1}}(\eta _{a}|g011\rangle +\eta
|e000\rangle ), \\
|1_{-}\rangle =& \frac{1}{\sqrt{2}}(|g100\rangle +\frac{\eta }{\beta _{1}}
|g011\rangle -\frac{\eta _{a}}{\beta _{1}}|e000\rangle ), \\
|1_{+}\rangle =& \frac{1}{\sqrt{2}}(|g100\rangle -\frac{\eta }{\beta _{1}}
|g011\rangle +\frac{\eta _{a}}{\beta _{1}}|e000\rangle ), \\
|2_{0}\rangle =& \frac{1}{A_{1}}(|g200\rangle +\frac{\eta _{a}^{2}-\eta ^{2}
}{\sqrt{2}\eta ^{2}}|g022\rangle +\frac{\sqrt{2}\eta _{a}}{\eta }
|e011\rangle ), \\
|2_{1-}\rangle =& \frac{1}{A_{2}}(d_{11}|g200\rangle +d_{12}|g111\rangle
+d_{13}|e100\rangle \\
& +d_{14}|g022\rangle +|e011\rangle ), \\
|2_{1+}\rangle =& \frac{1}{A_{2}}(d_{11}|g200\rangle -d_{12}|g111\rangle
-d_{13}|e100\rangle \\
& +d_{14}|g022\rangle +|e011\rangle ), \\
|2_{2-}\rangle =& \frac{1}{A_{3}}(d_{21}|g200\rangle +d_{22}|g111\rangle
+d_{23}|e100\rangle \\
& +d_{24}|g022\rangle +|e011\rangle ), \\
|2_{2+}\rangle =& \frac{1}{A_{3}}(d_{21}|g200\rangle -d_{22}|g111\rangle
-d_{23}|e100\rangle \\
& +d_{24}|g022\rangle +|e011\rangle ),
\end{split}
\label{dress}
\end{equation}
with the coefficients: $A_{1}=\frac{\sqrt{\beta _{1}^{4}+2\eta ^{4}}}{\sqrt{
2 }\eta ^{2}}$, $d_{11}=\frac{\beta _{1}^{2}(D-5\eta ^{2}-\eta _{a}^{2})}{
\sqrt{2}\eta _{a}\eta M_{1}}$, $d_{12}=\frac{\beta _{2}(-5\beta _{1}^{2}+D)}{
2\eta _{a}M_{1}}$, $d_{13}=\frac{\beta _{2}(\beta _{1}^{2}-D)}{2\eta M_{1}}$
, $d_{14}=\frac{\eta (D-5\beta _{1}^{2})}{\eta _{a}M_{1}}$, $d_{21}=-\frac{
\beta _{1}^{2}(D+5\eta ^{2}+\eta _{a}^{2})}{\sqrt{2}\eta _{a}\eta M_{2}}$, $
d_{22}=-\frac{\beta _{3}(5\beta _{1}^{2}+D)}{2\eta _{a}M_{2}}$, $d_{23}=
\frac{\beta _{2}(\beta _{1}^{2}+D)}{2\eta M_{2}}$, $d_{24}=-\frac{\eta
(D+5\beta _{1}^{2})}{\eta _{a}M_{2}}$, $M_{1}=3\beta _{1}^{2}-D$, $
M_{2}=3\beta _{1}^{2}+D$, and $A_{2(3)}=\sqrt{
|d_{1(2)1}|^{2}+|d_{1(2)2}|^{2}+|d_{1(2)3}|^{2}+|d_{1(2)4}|^{2}+1}$.

\section{The dynamic equation and steady states solution}

\label{solution_equation}
In this appendix, we derive probability amplitude
for a steady state. Substitute the $|\psi \rangle $ expressed by Eq.~ 
\eqref{eq:psi} into the Schr\"{o}dinger equation:
\begin{equation*}
i\frac{\partial }{\partial t}|\psi \rangle =H_{eff}|\psi \rangle ,
\end{equation*}
and we obtain the differential equations as%
\begin{eqnarray}  \label{Eq:append_2}
i\dot{C}_{g000} &=&0, \\
i\dot{C}_{g100} &=&\tilde{\Delta}C_{g100}-\eta C_{g011}+\eta _{a}C_{e000},
\notag \\
i\dot{C}_{g011} &=&-\eta C_{g100}+\tilde{\Delta}C_{g011},  \notag \\
i\dot{C}_{e000} &=&\eta _{a}C_{g100}+\tilde{\Delta}C_{e000}+\Omega
_{e}C_{g000},  \notag \\
i\dot{C}_{g200} &=&2\tilde{\Delta}C_{g200}-\sqrt{2}\eta C_{g111}+\sqrt{2}
\eta _{a}C_{e100},  \notag \\
i\dot{C}_{g111} &=&-\sqrt{2}\eta C_{g200}+2\tilde{\Delta}C_{g111}-2\eta
C_{g022}+\eta _{a}C_{e011},  \notag \\
i\dot{C}_{e100} &=&\Omega _{e}C_{g100}+\sqrt{2}\eta _{a}C_{g200}+2\tilde{
\Delta}C_{e100}-\eta C_{e011},  \notag \\
i\dot{C}_{g022} &=&-2\eta C_{g111}+2\tilde{\Delta}C_{g022},  \notag \\
i\dot{C}_{e011} &=&\Omega _{e}C_{g011}+\eta _{a}C_{g111}-\eta C_{e100}+2
\tilde{\Delta}C_{e011},  \notag
\end{eqnarray}%
where for simplicity, we set $\kappa _{+}=\kappa _{-}=\kappa _{e}=\kappa $ ,
$\tilde{\Delta}=\Delta -i\kappa $ and temporarily ignore the small
mechanical decay rate $\kappa _{b}\ll \kappa $, and the jumping from high
level to low level is ignored as it is done in Ref. \cite{PhysRevA.87.013839}.

The steady-state solution of Eq.~\eqref{Eq:append_2} is derived as
\begin{eqnarray}
C_{g000} &=&1 ,  \label{steadsolution} \\
C_{g100} &=&\frac{\eta _{a}\Omega _{e}}{\tilde{\Delta}^{2}-\eta
_{a}^{2}-\eta ^{2}} ,  \notag \\
C_{g011} &=&\frac{\eta \eta _{a}\Omega _{e}}{\tilde{\Delta}(\tilde{\Delta}
^{2}-\eta _{a}^{2}-\eta ^{2})},  \notag \\
C_{e000} &=&-\frac{(\tilde{\Delta}^{2}-\eta ^{2})\Omega _{e}}{\tilde{\Delta}%
( \tilde{\Delta}^{2}-\eta _{a}^{2}-\eta ^{2})},  \notag \\
C_{g111} &=&\frac{\eta _{a}^{2}\eta (5\tilde{\Delta}^{2}-\eta _{a}^{2}+\eta
^{2})\Omega_e^2}{\tilde{\Delta}B} ,  \notag \\
C_{g200} &=&\frac{\eta _{a}^{2}(4\tilde{\Delta}^{4}+\tilde{\Delta}^{2}(\eta
^{2}-\eta _{a}^{2})-2\eta ^{4})\Omega _{e}^{2}}{\sqrt{2}\tilde{\Delta}^{2}B},
\notag \\
C_{e100} &=&\frac{\eta _{a}(4\tilde{\Delta}^{4}-\tilde{\Delta}^{2}(\eta
_{a}^{2}+4\eta ^{2})+\eta ^{2}\eta _{a}^{2}-3\eta ^{4})\Omega _{e}^2}{\tilde{
\Delta}B},  \notag \\
C_{g022} &=&\frac{\eta ^{2}\eta _{a}^{2}(5\tilde{\Delta}^{2}-\eta
_{a}^{2}+\eta ^{2})\Omega _{e}^{2}}{\tilde{\Delta}^{2}B} ,  \notag \\
C_{e011} &=&-\frac{\eta _{a}\eta (6\tilde{\Delta}^{4}-\tilde{\Delta}
^{2}(\eta _{a}^{2}+9\eta ^{2})+2\eta ^{2}\eta _{a}^{2})\Omega _{e}^{2}}{%
\tilde{\Delta}^{2}B} ,  \notag
\end{eqnarray}%
where $B=\frac{1}{2}(\tilde{\Delta}^{2}-\beta _{1}^{2})(4\tilde{\Delta}%
^{2}-\beta _{2}^{2})(4\tilde{\Delta}^{2}-\beta _{3}^{2})$.

\section{The deduction of the relations between two bases}

\label{appendix_3} In this Appendix, we provide the certification of Eq.~%
\eqref{eq:relationship}. We define the Fock basis of the supermode $a_{\pm }$
as $|n_{+}n_{-}\rangle _{d}$ and the bare modes of $a$ and $m$ as $%
|nm\rangle $. For the supermodes, we have
\begin{equation}
\begin{split}
& a_{+}^{\dagger }|n_{+}n_{-}\rangle _{d}=\sqrt{n_{+}+1}|n_{+}+1n_{-}\rangle
_{d}, \\
& a_{+}|n_{+}n_{-}\rangle _{d}=\sqrt{n_{+}}|n_{+} -1 n_{-}\rangle _{d}, \\
& a_{-}^{\dagger }|n_{+}n_{-}\rangle _{d}=\sqrt{n_{-}+1}|n_{+}n_{-}+1\rangle
_{d}, \\
& a_{-}|n_{+}n_{-}\rangle _{d}=\sqrt{n_{-}}|n_{+}n_{-}-1\rangle _{d},
\end{split}%
\end{equation}%
Specifically, for $n_+,n_-=0$, we have the relation of the annihilation operator%
\begin{equation*}
a_{\pm }|00\rangle _{d}=0.
\end{equation*}%
Since $a_{\pm }=\frac{1}{\sqrt{2}}(a\pm m)$, we have $a|00\rangle _{d}=0$, $
m|00\rangle _{d}=0$. We expand the state $|00\rangle _{d}$ by using the bare
basis $|n,m\rangle $ of mode $a$ and $m$ as
\begin{equation}
\begin{split}
|00\rangle _{d}& =\sum_{n,m}C_{nm}|nm\rangle \\
& =C_{00}|00\rangle +C_{10}|10\rangle +C_{01}|01\rangle +\dots
\end{split}%
\end{equation}%
Thus
\begin{equation*}
C_{nm}=\langle nm|00\rangle _{d},
\end{equation*}%
for example $C_{10}=\langle 10|00\rangle _{d}=\langle 00|a|00\rangle _{d}=0$%
. Finally, we have
\begin{equation}
|00\rangle _{d}=|00\rangle .
\end{equation}%
In addition, we can write $a_{+}^{\dagger }|00\rangle _{d}=|10\rangle _{d}$,
i.e., $1/\sqrt{2}(a^{\dagger }+m^{\dagger })|00\rangle =1/\sqrt{2}(|10\rangle
+|01\rangle )$. Then, we can obtain
\begin{equation}
|10\rangle _{d}=\frac{1}{\sqrt{2}}(|10\rangle +|01\rangle ).
\label{eq:app3_1}
\end{equation}%
Similarly, we can have%
\begin{equation}
|01\rangle _{d}=\frac{1}{\sqrt{2}}(|10\rangle -|01\rangle ).
\label{eq:app3_2}
\end{equation}%
Taking action $a_{\pm }^{\dagger }=1/\sqrt{2}(a^{\dagger }\pm m^{\dagger })$
further on the right and left sides of Eq.~\eqref{eq:app3_1} and Eq.~ %
\eqref{eq:app3_2}, we can reach the other relations.

\bibliography{reference}

\begin{thebibliography}{54}%
\makeatletter
\providecommand \@ifxundefined [1]{%
 \@ifx{#1\undefined}
}%
\providecommand \@ifnum [1]{%
 \ifnum #1\expandafter \@firstoftwo
 \else \expandafter \@secondoftwo
 \fi
}%
\providecommand \@ifx [1]{%
 \ifx #1\expandafter \@firstoftwo
 \else \expandafter \@secondoftwo
 \fi
}%
\providecommand \natexlab [1]{#1}%
\providecommand \enquote  [1]{``#1''}%
\providecommand \bibnamefont  [1]{#1}%
\providecommand \bibfnamefont [1]{#1}%
\providecommand \citenamefont [1]{#1}%
\providecommand \href@noop [0]{\@secondoftwo}%
\providecommand \href [0]{\begingroup \@sanitize@url \@href}%
\providecommand \@href[1]{\@@startlink{#1}\@@href}%
\providecommand \@@href[1]{\endgroup#1\@@endlink}%
\providecommand \@sanitize@url [0]{\catcode `\\12\catcode `\$12\catcode
  `\&12\catcode `\#12\catcode `\^12\catcode `\_12\catcode `\%12\relax}%
\providecommand \@@startlink[1]{}%
\providecommand \@@endlink[0]{}%
\providecommand \url  [0]{\begingroup\@sanitize@url \@url }%
\providecommand \@url [1]{\endgroup\@href {#1}{\urlprefix }}%
\providecommand \urlprefix  [0]{URL }%
\providecommand \Eprint [0]{\href }%
\providecommand \doibase [0]{http://dx.doi.org/}%
\providecommand \selectlanguage [0]{\@gobble}%
\providecommand \bibinfo  [0]{\@secondoftwo}%
\providecommand \bibfield  [0]{\@secondoftwo}%
\providecommand \translation [1]{[#1]}%
\providecommand \BibitemOpen [0]{}%
\providecommand \bibitemStop [0]{}%
\providecommand \bibitemNoStop [0]{.\EOS\space}%
\providecommand \EOS [0]{\spacefactor3000\relax}%
\providecommand \BibitemShut  [1]{\csname bibitem#1\endcsname}%
\let\auto@bib@innerbib\@empty
%</preamble>
\bibitem [{\citenamefont {Birnbaum}\ \emph {et~al.}(2005)\citenamefont
  {Birnbaum}, \citenamefont {Boca}, \citenamefont {Miller}, \citenamefont
  {Boozer}, \citenamefont {Northup},\ and\ \citenamefont {Kimble}}]{Birnbaum}%
  \BibitemOpen
  \bibfield  {author} {\bibinfo {author} {\bibfnamefont {K.~M.}\ \bibnamefont
  {Birnbaum}}, \bibinfo {author} {\bibfnamefont {A.}~\bibnamefont {Boca}},
  \bibinfo {author} {\bibfnamefont {R.}~\bibnamefont {Miller}}, \bibinfo
  {author} {\bibfnamefont {A.~D.}\ \bibnamefont {Boozer}}, \bibinfo {author}
  {\bibfnamefont {T.~E.}\ \bibnamefont {Northup}}, \ and\ \bibinfo {author}
  {\bibfnamefont {H.~J.}\ \bibnamefont {Kimble}},\ }\href@noop {} {\bibfield
  {journal} {\bibinfo  {journal} {Nature}\ }\textbf {\bibinfo {volume} {436}},\
  \bibinfo {pages} {87} (\bibinfo {year} {2005})}\BibitemShut {NoStop}%
\bibitem [{\citenamefont {Imamo\ifmmode \bar{g}~\else \={g}\fi{}lu}\ \emph
  {et~al.}(1997)\citenamefont {Imamo\ifmmode \bar{g}~\else \={g}\fi{}lu},
  \citenamefont {Schmidt}, \citenamefont {Woods},\ and\ \citenamefont
  {Deutsch}}]{PhysRevLett.79.1467}%
  \BibitemOpen
  \bibfield  {author} {\bibinfo {author} {\bibfnamefont {A.}~\bibnamefont
  {Imamo\ifmmode \bar{g}~\else \={g}\fi{}lu}}, \bibinfo {author} {\bibfnamefont
  {H.}~\bibnamefont {Schmidt}}, \bibinfo {author} {\bibfnamefont
  {G.}~\bibnamefont {Woods}}, \ and\ \bibinfo {author} {\bibfnamefont
  {M.}~\bibnamefont {Deutsch}},\ }\href {\doibase 10.1103/PhysRevLett.79.1467}
  {\bibfield  {journal} {\bibinfo  {journal} {Phys. Rev. Lett.}\ }\textbf
  {\bibinfo {volume} {79}},\ \bibinfo {pages} {1467} (\bibinfo {year}
  {1997})}\BibitemShut {NoStop}%
\bibitem [{\citenamefont {Hacker}\ \emph {et~al.}(2016)\citenamefont {Hacker},
  \citenamefont {Welte}, \citenamefont {Rempe},\ and\ \citenamefont
  {Ritter}}]{hacker2016photon}%
  \BibitemOpen
  \bibfield  {author} {\bibinfo {author} {\bibfnamefont {B.}~\bibnamefont
  {Hacker}}, \bibinfo {author} {\bibfnamefont {S.}~\bibnamefont {Welte}},
  \bibinfo {author} {\bibfnamefont {G.}~\bibnamefont {Rempe}}, \ and\ \bibinfo
  {author} {\bibfnamefont {S.}~\bibnamefont {Ritter}},\ }\href@noop {}
  {\bibfield  {journal} {\bibinfo  {journal} {Nature}\ }\textbf {\bibinfo
  {volume} {536}},\ \bibinfo {pages} {193} (\bibinfo {year}
  {2016})}\BibitemShut {NoStop}%
\bibitem [{\citenamefont {Huang}\ \emph {et~al.}(2013)\citenamefont {Huang},
  \citenamefont {Liao},\ and\ \citenamefont {Sun}}]{PhysRevA.87.023822}%
  \BibitemOpen
  \bibfield  {author} {\bibinfo {author} {\bibfnamefont {J.-F.}\ \bibnamefont
  {Huang}}, \bibinfo {author} {\bibfnamefont {J.-Q.}\ \bibnamefont {Liao}}, \
  and\ \bibinfo {author} {\bibfnamefont {C.~P.}\ \bibnamefont {Sun}},\ }\href
  {\doibase 10.1103/PhysRevA.87.023822} {\bibfield  {journal} {\bibinfo
  {journal} {Phys. Rev. A}\ }\textbf {\bibinfo {volume} {87}},\ \bibinfo
  {pages} {023822} (\bibinfo {year} {2013})}\BibitemShut {NoStop}%
\bibitem [{\citenamefont {Deng}\ \emph {et~al.}(2015)\citenamefont {Deng},
  \citenamefont {Li},\ and\ \citenamefont {Qin}}]{PhysRevA.91.043831}%
  \BibitemOpen
  \bibfield  {author} {\bibinfo {author} {\bibfnamefont {W.-W.}\ \bibnamefont
  {Deng}}, \bibinfo {author} {\bibfnamefont {G.-X.}\ \bibnamefont {Li}}, \ and\
  \bibinfo {author} {\bibfnamefont {H.}~\bibnamefont {Qin}},\ }\href {\doibase
  10.1103/PhysRevA.91.043831} {\bibfield  {journal} {\bibinfo  {journal} {Phys.
  Rev. A}\ }\textbf {\bibinfo {volume} {91}},\ \bibinfo {pages} {043831}
  (\bibinfo {year} {2015})}\BibitemShut {NoStop}%
\bibitem [{\citenamefont {Zhu}\ \emph {et~al.}(2017)\citenamefont {Zhu},
  \citenamefont {Yang},\ and\ \citenamefont {Agarwal}}]{PhysRevA.95.063842}%
  \BibitemOpen
  \bibfield  {author} {\bibinfo {author} {\bibfnamefont {C.~J.}\ \bibnamefont
  {Zhu}}, \bibinfo {author} {\bibfnamefont {Y.~P.}\ \bibnamefont {Yang}}, \
  and\ \bibinfo {author} {\bibfnamefont {G.~S.}\ \bibnamefont {Agarwal}},\
  }\href {\doibase 10.1103/PhysRevA.95.063842} {\bibfield  {journal} {\bibinfo
  {journal} {Phys. Rev. A}\ }\textbf {\bibinfo {volume} {95}},\ \bibinfo
  {pages} {063842} (\bibinfo {year} {2017})}\BibitemShut {NoStop}%
\bibitem [{\citenamefont {Dayan}\ \emph {et~al.}(2008)\citenamefont {Dayan},
  \citenamefont {Parkins}, \citenamefont {Aoki}, \citenamefont {Ostby},
  \citenamefont {Vahala},\ and\ \citenamefont {Kimble}}]{Dayan1062}%
  \BibitemOpen
  \bibfield  {author} {\bibinfo {author} {\bibfnamefont {B.}~\bibnamefont
  {Dayan}}, \bibinfo {author} {\bibfnamefont {A.~S.}\ \bibnamefont {Parkins}},
  \bibinfo {author} {\bibfnamefont {T.}~\bibnamefont {Aoki}}, \bibinfo {author}
  {\bibfnamefont {E.~P.}\ \bibnamefont {Ostby}}, \bibinfo {author}
  {\bibfnamefont {K.~J.}\ \bibnamefont {Vahala}}, \ and\ \bibinfo {author}
  {\bibfnamefont {H.~J.}\ \bibnamefont {Kimble}},\ }\href@noop {} {\bibfield
  {journal} {\bibinfo  {journal} {Science}\ }\textbf {\bibinfo {volume}
  {319}},\ \bibinfo {pages} {1062} (\bibinfo {year} {2008})}\BibitemShut
  {NoStop}%
\bibitem [{\citenamefont {Snijders}\ \emph {et~al.}(2018)\citenamefont
  {Snijders}, \citenamefont {Frey}, \citenamefont {Norman}, \citenamefont
  {Flayac}, \citenamefont {Savona}, \citenamefont {Gossard}, \citenamefont
  {Bowers}, \citenamefont {van Exter}, \citenamefont {Bouwmeester},\ and\
  \citenamefont {L\"offler}}]{PhysRevLett.121.043601}%
  \BibitemOpen
  \bibfield  {author} {\bibinfo {author} {\bibfnamefont {H.~J.}\ \bibnamefont
  {Snijders}}, \bibinfo {author} {\bibfnamefont {J.~A.}\ \bibnamefont {Frey}},
  \bibinfo {author} {\bibfnamefont {J.}~\bibnamefont {Norman}}, \bibinfo
  {author} {\bibfnamefont {H.}~\bibnamefont {Flayac}}, \bibinfo {author}
  {\bibfnamefont {V.}~\bibnamefont {Savona}}, \bibinfo {author} {\bibfnamefont
  {A.~C.}\ \bibnamefont {Gossard}}, \bibinfo {author} {\bibfnamefont {J.~E.}\
  \bibnamefont {Bowers}}, \bibinfo {author} {\bibfnamefont {M.~P.}\
  \bibnamefont {van Exter}}, \bibinfo {author} {\bibfnamefont {D.}~\bibnamefont
  {Bouwmeester}}, \ and\ \bibinfo {author} {\bibfnamefont {W.}~\bibnamefont
  {L\"offler}},\ }\href {\doibase 10.1103/PhysRevLett.121.043601} {\bibfield
  {journal} {\bibinfo  {journal} {Phys. Rev. Lett.}\ }\textbf {\bibinfo
  {volume} {121}},\ \bibinfo {pages} {043601} (\bibinfo {year}
  {2018})}\BibitemShut {NoStop}%
\bibitem [{\citenamefont {Arcizet}\ \emph {et~al.}(2006)\citenamefont
  {Arcizet}, \citenamefont {Cohadon}, \citenamefont {Briant}, \citenamefont
  {Pinard}, \citenamefont {Heidmann}, \citenamefont {Mackowski}, \citenamefont
  {Michel}, \citenamefont {Pinard}, \citenamefont {Francais},\ and\
  \citenamefont {Rousseau}}]{PhysRevLett.97.133601}%
  \BibitemOpen
  \bibfield  {author} {\bibinfo {author} {\bibfnamefont {O.}~\bibnamefont
  {Arcizet}}, \bibinfo {author} {\bibfnamefont {P.-F.}\ \bibnamefont
  {Cohadon}}, \bibinfo {author} {\bibfnamefont {T.}~\bibnamefont {Briant}},
  \bibinfo {author} {\bibfnamefont {M.}~\bibnamefont {Pinard}}, \bibinfo
  {author} {\bibfnamefont {A.}~\bibnamefont {Heidmann}}, \bibinfo {author}
  {\bibfnamefont {J.-M.}\ \bibnamefont {Mackowski}}, \bibinfo {author}
  {\bibfnamefont {C.}~\bibnamefont {Michel}}, \bibinfo {author} {\bibfnamefont
  {L.}~\bibnamefont {Pinard}}, \bibinfo {author} {\bibfnamefont
  {O.}~\bibnamefont {Francais}}, \ and\ \bibinfo {author} {\bibfnamefont
  {L.}~\bibnamefont {Rousseau}},\ }\href {\doibase
  10.1103/PhysRevLett.97.133601} {\bibfield  {journal} {\bibinfo  {journal}
  {Phys. Rev. Lett.}\ }\textbf {\bibinfo {volume} {97}},\ \bibinfo {pages}
  {133601} (\bibinfo {year} {2006})}\BibitemShut {NoStop}%
\bibitem [{\citenamefont {Tsang}\ and\ \citenamefont
  {Caves}(2010)}]{PhysRevLett.105.123601}%
  \BibitemOpen
  \bibfield  {author} {\bibinfo {author} {\bibfnamefont {M.}~\bibnamefont
  {Tsang}}\ and\ \bibinfo {author} {\bibfnamefont {C.~M.}\ \bibnamefont
  {Caves}},\ }\href {\doibase 10.1103/PhysRevLett.105.123601} {\bibfield
  {journal} {\bibinfo  {journal} {Phys. Rev. Lett.}\ }\textbf {\bibinfo
  {volume} {105}},\ \bibinfo {pages} {123601} (\bibinfo {year}
  {2010})}\BibitemShut {NoStop}%
\bibitem [{\citenamefont {Nimmrichter}\ \emph {et~al.}(2014)\citenamefont
  {Nimmrichter}, \citenamefont {Hornberger},\ and\ \citenamefont
  {Hammerer}}]{PhysRevLett.113.020405}%
  \BibitemOpen
  \bibfield  {author} {\bibinfo {author} {\bibfnamefont {S.}~\bibnamefont
  {Nimmrichter}}, \bibinfo {author} {\bibfnamefont {K.}~\bibnamefont
  {Hornberger}}, \ and\ \bibinfo {author} {\bibfnamefont {K.}~\bibnamefont
  {Hammerer}},\ }\href {\doibase 10.1103/PhysRevLett.113.020405} {\bibfield
  {journal} {\bibinfo  {journal} {Phys. Rev. Lett.}\ }\textbf {\bibinfo
  {volume} {113}},\ \bibinfo {pages} {020405} (\bibinfo {year}
  {2014})}\BibitemShut {NoStop}%
\bibitem [{\citenamefont {Zhang}\ \emph
  {et~al.}(2017{\natexlab{a}})\citenamefont {Zhang}, \citenamefont {Han},
  \citenamefont {Xiong},\ and\ \citenamefont {Zhou}}]{Zhang_2017}%
  \BibitemOpen
  \bibfield  {author} {\bibinfo {author} {\bibfnamefont {W.-Z.}\ \bibnamefont
  {Zhang}}, \bibinfo {author} {\bibfnamefont {Y.}~\bibnamefont {Han}}, \bibinfo
  {author} {\bibfnamefont {B.}~\bibnamefont {Xiong}}, \ and\ \bibinfo {author}
  {\bibfnamefont {L.}~\bibnamefont {Zhou}},\ }\href {\doibase
  10.1088/1367-2630/aa68d9} {\bibfield  {journal} {\bibinfo  {journal} {New
  Journal of Physics}\ }\textbf {\bibinfo {volume} {19}},\ \bibinfo {pages}
  {083022} (\bibinfo {year} {2017}{\natexlab{a}})}\BibitemShut {NoStop}%
\bibitem [{\citenamefont {Liao}\ and\ \citenamefont
  {Tian}(2016)}]{PhysRevLett.116.163602}%
  \BibitemOpen
  \bibfield  {author} {\bibinfo {author} {\bibfnamefont {J.-Q.}\ \bibnamefont
  {Liao}}\ and\ \bibinfo {author} {\bibfnamefont {L.}~\bibnamefont {Tian}},\
  }\href {\doibase 10.1103/PhysRevLett.116.163602} {\bibfield  {journal}
  {\bibinfo  {journal} {Phys. Rev. Lett.}\ }\textbf {\bibinfo {volume} {116}},\
  \bibinfo {pages} {163602} (\bibinfo {year} {2016})}\BibitemShut {NoStop}%
\bibitem [{\citenamefont {Stannigel}\ \emph {et~al.}(2012)\citenamefont
  {Stannigel}, \citenamefont {Komar}, \citenamefont {Habraken}, \citenamefont
  {Bennett}, \citenamefont {Lukin}, \citenamefont {Zoller},\ and\ \citenamefont
  {Rabl}}]{PhysRevLett.109.013603}%
  \BibitemOpen
  \bibfield  {author} {\bibinfo {author} {\bibfnamefont {K.}~\bibnamefont
  {Stannigel}}, \bibinfo {author} {\bibfnamefont {P.}~\bibnamefont {Komar}},
  \bibinfo {author} {\bibfnamefont {S.~J.~M.}\ \bibnamefont {Habraken}},
  \bibinfo {author} {\bibfnamefont {S.~D.}\ \bibnamefont {Bennett}}, \bibinfo
  {author} {\bibfnamefont {M.~D.}\ \bibnamefont {Lukin}}, \bibinfo {author}
  {\bibfnamefont {P.}~\bibnamefont {Zoller}}, \ and\ \bibinfo {author}
  {\bibfnamefont {P.}~\bibnamefont {Rabl}},\ }\href {\doibase
  10.1103/PhysRevLett.109.013603} {\bibfield  {journal} {\bibinfo  {journal}
  {Phys. Rev. Lett.}\ }\textbf {\bibinfo {volume} {109}},\ \bibinfo {pages}
  {013603} (\bibinfo {year} {2012})}\BibitemShut {NoStop}%
\bibitem [{\citenamefont {Wang}\ and\ \citenamefont {Clerk}(2012)}]{Wang_2012}%
  \BibitemOpen
  \bibfield  {author} {\bibinfo {author} {\bibfnamefont {Y.-D.}\ \bibnamefont
  {Wang}}\ and\ \bibinfo {author} {\bibfnamefont {A.~A.}\ \bibnamefont
  {Clerk}},\ }\href {\doibase 10.1088/1367-2630/14/10/105010} {\bibfield
  {journal} {\bibinfo  {journal} {New Journal of Physics}\ }\textbf {\bibinfo
  {volume} {14}},\ \bibinfo {pages} {105010} (\bibinfo {year}
  {2012})}\BibitemShut {NoStop}%
\bibitem [{\citenamefont {Liu}\ \emph {et~al.}(2013)\citenamefont {Liu},
  \citenamefont {Xiao}, \citenamefont {Chen}, \citenamefont {Yu},\ and\
  \citenamefont {Gong}}]{PhysRevLett.111.083601}%
  \BibitemOpen
  \bibfield  {author} {\bibinfo {author} {\bibfnamefont {Y.-C.}\ \bibnamefont
  {Liu}}, \bibinfo {author} {\bibfnamefont {Y.-F.}\ \bibnamefont {Xiao}},
  \bibinfo {author} {\bibfnamefont {Y.-L.}\ \bibnamefont {Chen}}, \bibinfo
  {author} {\bibfnamefont {X.-C.}\ \bibnamefont {Yu}}, \ and\ \bibinfo {author}
  {\bibfnamefont {Q.}~\bibnamefont {Gong}},\ }\href {\doibase
  10.1103/PhysRevLett.111.083601} {\bibfield  {journal} {\bibinfo  {journal}
  {Phys. Rev. Lett.}\ }\textbf {\bibinfo {volume} {111}},\ \bibinfo {pages}
  {083601} (\bibinfo {year} {2013})}\BibitemShut {NoStop}%
\bibitem [{\citenamefont {Shen}\ \emph {et~al.}(2016)\citenamefont {Shen},
  \citenamefont {Zhang}, \citenamefont {Chen}, \citenamefont {Zou},
  \citenamefont {Xiao}, \citenamefont {Zou}, \citenamefont {Sun}, \citenamefont
  {Guo},\ and\ \citenamefont {Dong}}]{dongchunhua}%
  \BibitemOpen
  \bibfield  {author} {\bibinfo {author} {\bibfnamefont {Z.}~\bibnamefont
  {Shen}}, \bibinfo {author} {\bibfnamefont {Y.-L.}\ \bibnamefont {Zhang}},
  \bibinfo {author} {\bibfnamefont {Y.}~\bibnamefont {Chen}}, \bibinfo {author}
  {\bibfnamefont {C.-L.}\ \bibnamefont {Zou}}, \bibinfo {author} {\bibfnamefont
  {Y.-F.}\ \bibnamefont {Xiao}}, \bibinfo {author} {\bibfnamefont {X.-B.}\
  \bibnamefont {Zou}}, \bibinfo {author} {\bibfnamefont {F.-W.}\ \bibnamefont
  {Sun}}, \bibinfo {author} {\bibfnamefont {G.-C.}\ \bibnamefont {Guo}}, \ and\
  \bibinfo {author} {\bibfnamefont {C.-H.}\ \bibnamefont {Dong}},\ }\href@noop
  {} {\bibfield  {journal} {\bibinfo  {journal} {Nature Photonics}\ }\textbf
  {\bibinfo {volume} {10}},\ \bibinfo {pages} {657} (\bibinfo {year}
  {2016})}\BibitemShut {NoStop}%
\bibitem [{\citenamefont {Ridolfo}\ \emph {et~al.}(2012)\citenamefont
  {Ridolfo}, \citenamefont {Leib}, \citenamefont {Savasta},\ and\ \citenamefont
  {Hartmann}}]{PhysRevLett.109.193602}%
  \BibitemOpen
  \bibfield  {author} {\bibinfo {author} {\bibfnamefont {A.}~\bibnamefont
  {Ridolfo}}, \bibinfo {author} {\bibfnamefont {M.}~\bibnamefont {Leib}},
  \bibinfo {author} {\bibfnamefont {S.}~\bibnamefont {Savasta}}, \ and\
  \bibinfo {author} {\bibfnamefont {M.~J.}\ \bibnamefont {Hartmann}},\ }\href
  {\doibase 10.1103/PhysRevLett.109.193602} {\bibfield  {journal} {\bibinfo
  {journal} {Phys. Rev. Lett.}\ }\textbf {\bibinfo {volume} {109}},\ \bibinfo
  {pages} {193602} (\bibinfo {year} {2012})}\BibitemShut {NoStop}%
\bibitem [{\citenamefont {Zhang}\ \emph {et~al.}(2015)\citenamefont {Zhang},
  \citenamefont {Cheng}, \citenamefont {Liu},\ and\ \citenamefont
  {Zhou}}]{PhysRevA.91.063836}%
  \BibitemOpen
  \bibfield  {author} {\bibinfo {author} {\bibfnamefont {W.-Z.}\ \bibnamefont
  {Zhang}}, \bibinfo {author} {\bibfnamefont {J.}~\bibnamefont {Cheng}},
  \bibinfo {author} {\bibfnamefont {J.-Y.}\ \bibnamefont {Liu}}, \ and\
  \bibinfo {author} {\bibfnamefont {L.}~\bibnamefont {Zhou}},\ }\href {\doibase
  10.1103/PhysRevA.91.063836} {\bibfield  {journal} {\bibinfo  {journal} {Phys.
  Rev. A}\ }\textbf {\bibinfo {volume} {91}},\ \bibinfo {pages} {063836}
  (\bibinfo {year} {2015})}\BibitemShut {NoStop}%
\bibitem [{\citenamefont {Li}\ \emph {et~al.}(2016)\citenamefont {Li},
  \citenamefont {Zhang}, \citenamefont {Xiong},\ and\ \citenamefont
  {Zhou}}]{LiXun}%
  \BibitemOpen
  \bibfield  {author} {\bibinfo {author} {\bibfnamefont {X.}~\bibnamefont
  {Li}}, \bibinfo {author} {\bibfnamefont {W.-Z.}\ \bibnamefont {Zhang}},
  \bibinfo {author} {\bibfnamefont {B.}~\bibnamefont {Xiong}}, \ and\ \bibinfo
  {author} {\bibfnamefont {L.}~\bibnamefont {Zhou}},\ }\href@noop {} {\bibfield
   {journal} {\bibinfo  {journal} {Scientific Reports}\ }\textbf {\bibinfo
  {volume} {6}},\ \bibinfo {pages} {39343} (\bibinfo {year}
  {2016})}\BibitemShut {NoStop}%
\bibitem [{\citenamefont {Rabl}(2011)}]{PhysRevLett.107.063601}%
  \BibitemOpen
  \bibfield  {author} {\bibinfo {author} {\bibfnamefont {P.}~\bibnamefont
  {Rabl}},\ }\href {\doibase 10.1103/PhysRevLett.107.063601} {\bibfield
  {journal} {\bibinfo  {journal} {Phys. Rev. Lett.}\ }\textbf {\bibinfo
  {volume} {107}},\ \bibinfo {pages} {063601} (\bibinfo {year}
  {2011})}\BibitemShut {NoStop}%
\bibitem [{\citenamefont {Liao}\ and\ \citenamefont
  {Nori}(2013)}]{PhysRevA.88.023853}%
  \BibitemOpen
  \bibfield  {author} {\bibinfo {author} {\bibfnamefont {J.-Q.}\ \bibnamefont
  {Liao}}\ and\ \bibinfo {author} {\bibfnamefont {F.}~\bibnamefont {Nori}},\
  }\href {\doibase 10.1103/PhysRevA.88.023853} {\bibfield  {journal} {\bibinfo
  {journal} {Phys. Rev. A}\ }\textbf {\bibinfo {volume} {88}},\ \bibinfo
  {pages} {023853} (\bibinfo {year} {2013})}\BibitemShut {NoStop}%
\bibitem [{\citenamefont {Zhou}\ \emph {et~al.}(2013)\citenamefont {Zhou},
  \citenamefont {Cheng}, \citenamefont {Han},\ and\ \citenamefont
  {Zhang}}]{PhysRevA.88.063854}%
  \BibitemOpen
  \bibfield  {author} {\bibinfo {author} {\bibfnamefont {L.}~\bibnamefont
  {Zhou}}, \bibinfo {author} {\bibfnamefont {J.}~\bibnamefont {Cheng}},
  \bibinfo {author} {\bibfnamefont {Y.}~\bibnamefont {Han}}, \ and\ \bibinfo
  {author} {\bibfnamefont {W.}~\bibnamefont {Zhang}},\ }\href {\doibase
  10.1103/PhysRevA.88.063854} {\bibfield  {journal} {\bibinfo  {journal} {Phys.
  Rev. A}\ }\textbf {\bibinfo {volume} {88}},\ \bibinfo {pages} {063854}
  (\bibinfo {year} {2013})}\BibitemShut {NoStop}%
\bibitem [{\citenamefont {Ludwig}\ \emph {et~al.}(2012)\citenamefont {Ludwig},
  \citenamefont {Safavi-Naeini}, \citenamefont {Painter},\ and\ \citenamefont
  {Marquardt}}]{PhysRevLett.109.063601}%
  \BibitemOpen
  \bibfield  {author} {\bibinfo {author} {\bibfnamefont {M.}~\bibnamefont
  {Ludwig}}, \bibinfo {author} {\bibfnamefont {A.~H.}\ \bibnamefont
  {Safavi-Naeini}}, \bibinfo {author} {\bibfnamefont {O.}~\bibnamefont
  {Painter}}, \ and\ \bibinfo {author} {\bibfnamefont {F.}~\bibnamefont
  {Marquardt}},\ }\href {\doibase 10.1103/PhysRevLett.109.063601} {\bibfield
  {journal} {\bibinfo  {journal} {Phys. Rev. Lett.}\ }\textbf {\bibinfo
  {volume} {109}},\ \bibinfo {pages} {063601} (\bibinfo {year}
  {2012})}\BibitemShut {NoStop}%
\bibitem [{\citenamefont {K\'om\'ar}\ \emph {et~al.}(2013)\citenamefont
  {K\'om\'ar}, \citenamefont {Bennett}, \citenamefont {Stannigel},
  \citenamefont {Habraken}, \citenamefont {Rabl}, \citenamefont {Zoller},\ and\
  \citenamefont {Lukin}}]{PhysRevA.87.013839}%
  \BibitemOpen
  \bibfield  {author} {\bibinfo {author} {\bibfnamefont {P.}~\bibnamefont
  {K\'om\'ar}}, \bibinfo {author} {\bibfnamefont {S.~D.}\ \bibnamefont
  {Bennett}}, \bibinfo {author} {\bibfnamefont {K.}~\bibnamefont {Stannigel}},
  \bibinfo {author} {\bibfnamefont {S.~J.~M.}\ \bibnamefont {Habraken}},
  \bibinfo {author} {\bibfnamefont {P.}~\bibnamefont {Rabl}}, \bibinfo {author}
  {\bibfnamefont {P.}~\bibnamefont {Zoller}}, \ and\ \bibinfo {author}
  {\bibfnamefont {M.~D.}\ \bibnamefont {Lukin}},\ }\href {\doibase
  10.1103/PhysRevA.87.013839} {\bibfield  {journal} {\bibinfo  {journal} {Phys.
  Rev. A}\ }\textbf {\bibinfo {volume} {87}},\ \bibinfo {pages} {013839}
  (\bibinfo {year} {2013})}\BibitemShut {NoStop}%
\bibitem [{\citenamefont {Vaneph}\ \emph {et~al.}(2018)\citenamefont {Vaneph},
  \citenamefont {Morvan}, \citenamefont {Aiello}, \citenamefont {F\'echant},
  \citenamefont {Aprili}, \citenamefont {Gabelli},\ and\ \citenamefont
  {Est\`eve}}]{PhysRevLett.121.043602}%
  \BibitemOpen
  \bibfield  {author} {\bibinfo {author} {\bibfnamefont {C.}~\bibnamefont
  {Vaneph}}, \bibinfo {author} {\bibfnamefont {A.}~\bibnamefont {Morvan}},
  \bibinfo {author} {\bibfnamefont {G.}~\bibnamefont {Aiello}}, \bibinfo
  {author} {\bibfnamefont {M.}~\bibnamefont {F\'echant}}, \bibinfo {author}
  {\bibfnamefont {M.}~\bibnamefont {Aprili}}, \bibinfo {author} {\bibfnamefont
  {J.}~\bibnamefont {Gabelli}}, \ and\ \bibinfo {author} {\bibfnamefont
  {J.}~\bibnamefont {Est\`eve}},\ }\href {\doibase
  10.1103/PhysRevLett.121.043602} {\bibfield  {journal} {\bibinfo  {journal}
  {Phys. Rev. Lett.}\ }\textbf {\bibinfo {volume} {121}},\ \bibinfo {pages}
  {043602} (\bibinfo {year} {2018})}\BibitemShut {NoStop}%
\bibitem [{\citenamefont {Zhang}\ \emph {et~al.}(2014)\citenamefont {Zhang},
  \citenamefont {Zou}, \citenamefont {Jiang},\ and\ \citenamefont
  {Tang}}]{PhysRevLett.113.156401}%
  \BibitemOpen
  \bibfield  {author} {\bibinfo {author} {\bibfnamefont {X.}~\bibnamefont
  {Zhang}}, \bibinfo {author} {\bibfnamefont {C.-L.}\ \bibnamefont {Zou}},
  \bibinfo {author} {\bibfnamefont {L.}~\bibnamefont {Jiang}}, \ and\ \bibinfo
  {author} {\bibfnamefont {H.~X.}\ \bibnamefont {Tang}},\ }\href {\doibase
  10.1103/PhysRevLett.113.156401} {\bibfield  {journal} {\bibinfo  {journal}
  {Phys. Rev. Lett.}\ }\textbf {\bibinfo {volume} {113}},\ \bibinfo {pages}
  {156401} (\bibinfo {year} {2014})}\BibitemShut {NoStop}%
\bibitem [{\citenamefont {Bourhill}\ \emph {et~al.}(2016)\citenamefont
  {Bourhill}, \citenamefont {Kostylev}, \citenamefont {Goryachev},
  \citenamefont {Creedon},\ and\ \citenamefont {Tobar}}]{PhysRevB.93.144420}%
  \BibitemOpen
  \bibfield  {author} {\bibinfo {author} {\bibfnamefont {J.}~\bibnamefont
  {Bourhill}}, \bibinfo {author} {\bibfnamefont {N.}~\bibnamefont {Kostylev}},
  \bibinfo {author} {\bibfnamefont {M.}~\bibnamefont {Goryachev}}, \bibinfo
  {author} {\bibfnamefont {D.~L.}\ \bibnamefont {Creedon}}, \ and\ \bibinfo
  {author} {\bibfnamefont {M.~E.}\ \bibnamefont {Tobar}},\ }\href {\doibase
  10.1103/PhysRevB.93.144420} {\bibfield  {journal} {\bibinfo  {journal} {Phys.
  Rev. B}\ }\textbf {\bibinfo {volume} {93}},\ \bibinfo {pages} {144420}
  (\bibinfo {year} {2016})}\BibitemShut {NoStop}%
\bibitem [{\citenamefont {Goryachev}\ \emph {et~al.}(2014)\citenamefont
  {Goryachev}, \citenamefont {Farr}, \citenamefont {Creedon}, \citenamefont
  {Fan}, \citenamefont {Kostylev},\ and\ \citenamefont
  {Tobar}}]{PhysRevApplied.2.054002}%
  \BibitemOpen
  \bibfield  {author} {\bibinfo {author} {\bibfnamefont {M.}~\bibnamefont
  {Goryachev}}, \bibinfo {author} {\bibfnamefont {W.~G.}\ \bibnamefont {Farr}},
  \bibinfo {author} {\bibfnamefont {D.~L.}\ \bibnamefont {Creedon}}, \bibinfo
  {author} {\bibfnamefont {Y.}~\bibnamefont {Fan}}, \bibinfo {author}
  {\bibfnamefont {M.}~\bibnamefont {Kostylev}}, \ and\ \bibinfo {author}
  {\bibfnamefont {M.~E.}\ \bibnamefont {Tobar}},\ }\href {\doibase
  10.1103/PhysRevApplied.2.054002} {\bibfield  {journal} {\bibinfo  {journal}
  {Phys. Rev. Applied}\ }\textbf {\bibinfo {volume} {2}},\ \bibinfo {pages}
  {054002} (\bibinfo {year} {2014})}\BibitemShut {NoStop}%
\bibitem [{\citenamefont {Zhang}\ \emph
  {et~al.}(2016{\natexlab{a}})\citenamefont {Zhang}, \citenamefont {Zhu},
  \citenamefont {Zou},\ and\ \citenamefont {Tang}}]{PhysRevLett.117.123605}%
  \BibitemOpen
  \bibfield  {author} {\bibinfo {author} {\bibfnamefont {X.}~\bibnamefont
  {Zhang}}, \bibinfo {author} {\bibfnamefont {N.}~\bibnamefont {Zhu}}, \bibinfo
  {author} {\bibfnamefont {C.-L.}\ \bibnamefont {Zou}}, \ and\ \bibinfo
  {author} {\bibfnamefont {H.~X.}\ \bibnamefont {Tang}},\ }\href {\doibase
  10.1103/PhysRevLett.117.123605} {\bibfield  {journal} {\bibinfo  {journal}
  {Phys. Rev. Lett.}\ }\textbf {\bibinfo {volume} {117}},\ \bibinfo {pages}
  {123605} (\bibinfo {year} {2016}{\natexlab{a}})}\BibitemShut {NoStop}%
\bibitem [{\citenamefont {Haigh}\ \emph {et~al.}(2016)\citenamefont {Haigh},
  \citenamefont {Nunnenkamp}, \citenamefont {Ramsay},\ and\ \citenamefont
  {Ferguson}}]{PhysRevLett.117.133602}%
  \BibitemOpen
  \bibfield  {author} {\bibinfo {author} {\bibfnamefont {J.~A.}\ \bibnamefont
  {Haigh}}, \bibinfo {author} {\bibfnamefont {A.}~\bibnamefont {Nunnenkamp}},
  \bibinfo {author} {\bibfnamefont {A.~J.}\ \bibnamefont {Ramsay}}, \ and\
  \bibinfo {author} {\bibfnamefont {A.~J.}\ \bibnamefont {Ferguson}},\ }\href
  {\doibase 10.1103/PhysRevLett.117.133602} {\bibfield  {journal} {\bibinfo
  {journal} {Phys. Rev. Lett.}\ }\textbf {\bibinfo {volume} {117}},\ \bibinfo
  {pages} {133602} (\bibinfo {year} {2016})}\BibitemShut {NoStop}%
\bibitem [{\citenamefont {Osada}\ \emph {et~al.}(2016)\citenamefont {Osada},
  \citenamefont {Hisatomi}, \citenamefont {Noguchi}, \citenamefont {Tabuchi},
  \citenamefont {Yamazaki}, \citenamefont {Usami}, \citenamefont {Sadgrove},
  \citenamefont {Yalla}, \citenamefont {Nomura},\ and\ \citenamefont
  {Nakamura}}]{PhysRevLett.116.223601}%
  \BibitemOpen
  \bibfield  {author} {\bibinfo {author} {\bibfnamefont {A.}~\bibnamefont
  {Osada}}, \bibinfo {author} {\bibfnamefont {R.}~\bibnamefont {Hisatomi}},
  \bibinfo {author} {\bibfnamefont {A.}~\bibnamefont {Noguchi}}, \bibinfo
  {author} {\bibfnamefont {Y.}~\bibnamefont {Tabuchi}}, \bibinfo {author}
  {\bibfnamefont {R.}~\bibnamefont {Yamazaki}}, \bibinfo {author}
  {\bibfnamefont {K.}~\bibnamefont {Usami}}, \bibinfo {author} {\bibfnamefont
  {M.}~\bibnamefont {Sadgrove}}, \bibinfo {author} {\bibfnamefont
  {R.}~\bibnamefont {Yalla}}, \bibinfo {author} {\bibfnamefont
  {M.}~\bibnamefont {Nomura}}, \ and\ \bibinfo {author} {\bibfnamefont
  {Y.}~\bibnamefont {Nakamura}},\ }\href {\doibase
  10.1103/PhysRevLett.116.223601} {\bibfield  {journal} {\bibinfo  {journal}
  {Phys. Rev. Lett.}\ }\textbf {\bibinfo {volume} {116}},\ \bibinfo {pages}
  {223601} (\bibinfo {year} {2016})}\BibitemShut {NoStop}%
\bibitem [{\citenamefont {Niemczyk}\ \emph {et~al.}(2010)\citenamefont
  {Niemczyk}, \citenamefont {Deppe}, \citenamefont {Huebl}, \citenamefont
  {Menzel}, \citenamefont {Hocke}, \citenamefont {Schwarz}, \citenamefont
  {Garcia-Ripoll}, \citenamefont {Zueco}, \citenamefont {H{\"u}mmer},
  \citenamefont {Solano} \emph {et~al.}}]{niemczyk2010circuit}%
  \BibitemOpen
  \bibfield  {author} {\bibinfo {author} {\bibfnamefont {T.}~\bibnamefont
  {Niemczyk}}, \bibinfo {author} {\bibfnamefont {F.}~\bibnamefont {Deppe}},
  \bibinfo {author} {\bibfnamefont {H.}~\bibnamefont {Huebl}}, \bibinfo
  {author} {\bibfnamefont {E.}~\bibnamefont {Menzel}}, \bibinfo {author}
  {\bibfnamefont {F.}~\bibnamefont {Hocke}}, \bibinfo {author} {\bibfnamefont
  {M.}~\bibnamefont {Schwarz}}, \bibinfo {author} {\bibfnamefont
  {J.}~\bibnamefont {Garcia-Ripoll}}, \bibinfo {author} {\bibfnamefont
  {D.}~\bibnamefont {Zueco}}, \bibinfo {author} {\bibfnamefont
  {T.}~\bibnamefont {H{\"u}mmer}}, \bibinfo {author} {\bibfnamefont
  {E.}~\bibnamefont {Solano}},  \emph {et~al.},\ }\href@noop {} {\bibfield
  {journal} {\bibinfo  {journal} {Nature Physics}\ }\textbf {\bibinfo {volume}
  {6}},\ \bibinfo {pages} {772} (\bibinfo {year} {2010})}\BibitemShut {NoStop}%
\bibitem [{\citenamefont {Li}\ \emph {et~al.}(2018)\citenamefont {Li},
  \citenamefont {Zhu},\ and\ \citenamefont {Agarwal}}]{PhysRevLett.121.203601}%
  \BibitemOpen
  \bibfield  {author} {\bibinfo {author} {\bibfnamefont {J.}~\bibnamefont
  {Li}}, \bibinfo {author} {\bibfnamefont {S.-Y.}\ \bibnamefont {Zhu}}, \ and\
  \bibinfo {author} {\bibfnamefont {G.~S.}\ \bibnamefont {Agarwal}},\ }\href
  {\doibase 10.1103/PhysRevLett.121.203601} {\bibfield  {journal} {\bibinfo
  {journal} {Phys. Rev. Lett.}\ }\textbf {\bibinfo {volume} {121}},\ \bibinfo
  {pages} {203601} (\bibinfo {year} {2018})}\BibitemShut {NoStop}%
\bibitem [{\citenamefont {Zhang}\ \emph
  {et~al.}(2016{\natexlab{b}})\citenamefont {Zhang}, \citenamefont {Zou},
  \citenamefont {Jiang},\ and\ \citenamefont {Tang}}]{zhang2016cavity}%
  \BibitemOpen
  \bibfield  {author} {\bibinfo {author} {\bibfnamefont {X.}~\bibnamefont
  {Zhang}}, \bibinfo {author} {\bibfnamefont {C.-L.}\ \bibnamefont {Zou}},
  \bibinfo {author} {\bibfnamefont {L.}~\bibnamefont {Jiang}}, \ and\ \bibinfo
  {author} {\bibfnamefont {H.~X.}\ \bibnamefont {Tang}},\ }\href@noop {}
  {\bibfield  {journal} {\bibinfo  {journal} {Sci. Adv.}\ }\textbf {\bibinfo
  {volume} {2}},\ \bibinfo {pages} {e1501286} (\bibinfo {year}
  {2016}{\natexlab{b}})}\BibitemShut {NoStop}%
\bibitem [{\citenamefont {Gao}\ \emph {et~al.}(2017)\citenamefont {Gao},
  \citenamefont {Cao}, \citenamefont {Wang}, \citenamefont {Zhang},\ and\
  \citenamefont {Wang}}]{PhysRevA.96.023826}%
  \BibitemOpen
  \bibfield  {author} {\bibinfo {author} {\bibfnamefont {Y.-P.}\ \bibnamefont
  {Gao}}, \bibinfo {author} {\bibfnamefont {C.}~\bibnamefont {Cao}}, \bibinfo
  {author} {\bibfnamefont {T.-J.}\ \bibnamefont {Wang}}, \bibinfo {author}
  {\bibfnamefont {Y.}~\bibnamefont {Zhang}}, \ and\ \bibinfo {author}
  {\bibfnamefont {C.}~\bibnamefont {Wang}},\ }\href {\doibase
  10.1103/PhysRevA.96.023826} {\bibfield  {journal} {\bibinfo  {journal} {Phys.
  Rev. A}\ }\textbf {\bibinfo {volume} {96}},\ \bibinfo {pages} {023826}
  (\bibinfo {year} {2017})}\BibitemShut {NoStop}%
\bibitem [{\citenamefont {Gao}\ \emph {et~al.}(2019)\citenamefont {Gao},
  \citenamefont {Liu}, \citenamefont {Wang}, \citenamefont {Cao},\ and\
  \citenamefont {Wang}}]{PhysRevA.100.043831}%
  \BibitemOpen
  \bibfield  {author} {\bibinfo {author} {\bibfnamefont {Y.-P.}\ \bibnamefont
  {Gao}}, \bibinfo {author} {\bibfnamefont {X.-F.}\ \bibnamefont {Liu}},
  \bibinfo {author} {\bibfnamefont {T.-J.}\ \bibnamefont {Wang}}, \bibinfo
  {author} {\bibfnamefont {C.}~\bibnamefont {Cao}}, \ and\ \bibinfo {author}
  {\bibfnamefont {C.}~\bibnamefont {Wang}},\ }\href {\doibase
  10.1103/PhysRevA.100.043831} {\bibfield  {journal} {\bibinfo  {journal}
  {Phys. Rev. A}\ }\textbf {\bibinfo {volume} {100}},\ \bibinfo {pages}
  {043831} (\bibinfo {year} {2019})}\BibitemShut {NoStop}%
\bibitem [{\citenamefont {Flayac}\ and\ \citenamefont
  {Savona}(2017)}]{PhysRevA.96.053810}%
  \BibitemOpen
  \bibfield  {author} {\bibinfo {author} {\bibfnamefont {H.}~\bibnamefont
  {Flayac}}\ and\ \bibinfo {author} {\bibfnamefont {V.}~\bibnamefont
  {Savona}},\ }\href {\doibase 10.1103/PhysRevA.96.053810} {\bibfield
  {journal} {\bibinfo  {journal} {Phys. Rev. A}\ }\textbf {\bibinfo {volume}
  {96}},\ \bibinfo {pages} {053810} (\bibinfo {year} {2017})}\BibitemShut
  {NoStop}%
\bibitem [{\citenamefont {Shen}\ \emph {et~al.}(2014)\citenamefont {Shen},
  \citenamefont {Zhou},\ and\ \citenamefont {Yi}}]{PhysRevA.90.023849}%
  \BibitemOpen
  \bibfield  {author} {\bibinfo {author} {\bibfnamefont {H.~Z.}\ \bibnamefont
  {Shen}}, \bibinfo {author} {\bibfnamefont {Y.~H.}\ \bibnamefont {Zhou}}, \
  and\ \bibinfo {author} {\bibfnamefont {X.~X.}\ \bibnamefont {Yi}},\ }\href
  {\doibase 10.1103/PhysRevA.90.023849} {\bibfield  {journal} {\bibinfo
  {journal} {Phys. Rev. A}\ }\textbf {\bibinfo {volume} {90}},\ \bibinfo
  {pages} {023849} (\bibinfo {year} {2014})}\BibitemShut {NoStop}%
\bibitem [{\citenamefont {Majumdar}\ and\ \citenamefont
  {Gerace}(2013)}]{PhysRevB.87.235319}%
  \BibitemOpen
  \bibfield  {author} {\bibinfo {author} {\bibfnamefont {A.}~\bibnamefont
  {Majumdar}}\ and\ \bibinfo {author} {\bibfnamefont {D.}~\bibnamefont
  {Gerace}},\ }\href {\doibase 10.1103/PhysRevB.87.235319} {\bibfield
  {journal} {\bibinfo  {journal} {Phys. Rev. B}\ }\textbf {\bibinfo {volume}
  {87}},\ \bibinfo {pages} {235319} (\bibinfo {year} {2013})}\BibitemShut
  {NoStop}%
\bibitem [{\citenamefont {Zou}\ \emph {et~al.}(2019)\citenamefont {Zou},
  \citenamefont {Fan}, \citenamefont {Huang},\ and\ \citenamefont
  {Liao}}]{PhysRevA.99.043837}%
  \BibitemOpen
  \bibfield  {author} {\bibinfo {author} {\bibfnamefont {F.}~\bibnamefont
  {Zou}}, \bibinfo {author} {\bibfnamefont {L.-B.}\ \bibnamefont {Fan}},
  \bibinfo {author} {\bibfnamefont {J.-F.}\ \bibnamefont {Huang}}, \ and\
  \bibinfo {author} {\bibfnamefont {J.-Q.}\ \bibnamefont {Liao}},\ }\href
  {\doibase 10.1103/PhysRevA.99.043837} {\bibfield  {journal} {\bibinfo
  {journal} {Phys. Rev. A}\ }\textbf {\bibinfo {volume} {99}},\ \bibinfo
  {pages} {043837} (\bibinfo {year} {2019})}\BibitemShut {NoStop}%
\bibitem [{\citenamefont {Liu}\ \emph {et~al.}(2019)\citenamefont {Liu},
  \citenamefont {Xiong},\ and\ \citenamefont {Wu}}]{PhysRevB.100.134421}%
  \BibitemOpen
  \bibfield  {author} {\bibinfo {author} {\bibfnamefont {Z.-X.}\ \bibnamefont
  {Liu}}, \bibinfo {author} {\bibfnamefont {H.}~\bibnamefont {Xiong}}, \ and\
  \bibinfo {author} {\bibfnamefont {Y.}~\bibnamefont {Wu}},\ }\href {\doibase
  10.1103/PhysRevB.100.134421} {\bibfield  {journal} {\bibinfo  {journal}
  {Phys. Rev. B}\ }\textbf {\bibinfo {volume} {100}},\ \bibinfo {pages}
  {134421} (\bibinfo {year} {2019})}\BibitemShut {NoStop}%
\bibitem [{\citenamefont {Tashima}\ \emph {et~al.}(2019)\citenamefont
  {Tashima}, \citenamefont {Morishita},\ and\ \citenamefont
  {Mizuochi}}]{PhysRevA.100.023801}%
  \BibitemOpen
  \bibfield  {author} {\bibinfo {author} {\bibfnamefont {T.}~\bibnamefont
  {Tashima}}, \bibinfo {author} {\bibfnamefont {H.}~\bibnamefont {Morishita}},
  \ and\ \bibinfo {author} {\bibfnamefont {N.}~\bibnamefont {Mizuochi}},\
  }\href {\doibase 10.1103/PhysRevA.100.023801} {\bibfield  {journal} {\bibinfo
   {journal} {Phys. Rev. A}\ }\textbf {\bibinfo {volume} {100}},\ \bibinfo
  {pages} {023801} (\bibinfo {year} {2019})}\BibitemShut {NoStop}%
\bibitem [{\citenamefont {Huebl}\ \emph {et~al.}(2013)\citenamefont {Huebl},
  \citenamefont {Zollitsch}, \citenamefont {Lotze}, \citenamefont {Hocke},
  \citenamefont {Greifenstein}, \citenamefont {Marx}, \citenamefont {Gross},\
  and\ \citenamefont {Goennenwein}}]{PhysRevLett.111.127003}%
  \BibitemOpen
  \bibfield  {author} {\bibinfo {author} {\bibfnamefont {H.}~\bibnamefont
  {Huebl}}, \bibinfo {author} {\bibfnamefont {C.~W.}\ \bibnamefont
  {Zollitsch}}, \bibinfo {author} {\bibfnamefont {J.}~\bibnamefont {Lotze}},
  \bibinfo {author} {\bibfnamefont {F.}~\bibnamefont {Hocke}}, \bibinfo
  {author} {\bibfnamefont {M.}~\bibnamefont {Greifenstein}}, \bibinfo {author}
  {\bibfnamefont {A.}~\bibnamefont {Marx}}, \bibinfo {author} {\bibfnamefont
  {R.}~\bibnamefont {Gross}}, \ and\ \bibinfo {author} {\bibfnamefont
  {S.~T.~B.}\ \bibnamefont {Goennenwein}},\ }\href {\doibase
  10.1103/PhysRevLett.111.127003} {\bibfield  {journal} {\bibinfo  {journal}
  {Phys. Rev. Lett.}\ }\textbf {\bibinfo {volume} {111}},\ \bibinfo {pages}
  {127003} (\bibinfo {year} {2013})}\BibitemShut {NoStop}%
\bibitem [{\citenamefont {Wang}\ \emph {et~al.}(2018)\citenamefont {Wang},
  \citenamefont {Zhang}, \citenamefont {Zhang}, \citenamefont {Li},
  \citenamefont {Hu},\ and\ \citenamefont {You}}]{PhysRevLett.120.057202}%
  \BibitemOpen
  \bibfield  {author} {\bibinfo {author} {\bibfnamefont {Y.-P.}\ \bibnamefont
  {Wang}}, \bibinfo {author} {\bibfnamefont {G.-Q.}\ \bibnamefont {Zhang}},
  \bibinfo {author} {\bibfnamefont {D.}~\bibnamefont {Zhang}}, \bibinfo
  {author} {\bibfnamefont {T.-F.}\ \bibnamefont {Li}}, \bibinfo {author}
  {\bibfnamefont {C.-M.}\ \bibnamefont {Hu}}, \ and\ \bibinfo {author}
  {\bibfnamefont {J.~Q.}\ \bibnamefont {You}},\ }\href {\doibase
  10.1103/PhysRevLett.120.057202} {\bibfield  {journal} {\bibinfo  {journal}
  {Phys. Rev. Lett.}\ }\textbf {\bibinfo {volume} {120}},\ \bibinfo {pages}
  {057202} (\bibinfo {year} {2018})}\BibitemShut {NoStop}%
\bibitem [{\citenamefont {Hyde}\ \emph {et~al.}(2018)\citenamefont {Hyde},
  \citenamefont {Yao}, \citenamefont {Gui}, \citenamefont {Zhang},
  \citenamefont {You},\ and\ \citenamefont {Hu}}]{PhysRevB.98.174423}%
  \BibitemOpen
  \bibfield  {author} {\bibinfo {author} {\bibfnamefont {P.}~\bibnamefont
  {Hyde}}, \bibinfo {author} {\bibfnamefont {B.~M.}\ \bibnamefont {Yao}},
  \bibinfo {author} {\bibfnamefont {Y.~S.}\ \bibnamefont {Gui}}, \bibinfo
  {author} {\bibfnamefont {G.-Q.}\ \bibnamefont {Zhang}}, \bibinfo {author}
  {\bibfnamefont {J.~Q.}\ \bibnamefont {You}}, \ and\ \bibinfo {author}
  {\bibfnamefont {C.-M.}\ \bibnamefont {Hu}},\ }\href {\doibase
  10.1103/PhysRevB.98.174423} {\bibfield  {journal} {\bibinfo  {journal} {Phys.
  Rev. B}\ }\textbf {\bibinfo {volume} {98}},\ \bibinfo {pages} {174423}
  (\bibinfo {year} {2018})}\BibitemShut {NoStop}%
\bibitem [{\citenamefont {Zhang}\ \emph
  {et~al.}(2017{\natexlab{b}})\citenamefont {Zhang}, \citenamefont {Luo},
  \citenamefont {Wang}, \citenamefont {Li},\ and\ \citenamefont
  {You}}]{zhang2017observation}%
  \BibitemOpen
  \bibfield  {author} {\bibinfo {author} {\bibfnamefont {D.}~\bibnamefont
  {Zhang}}, \bibinfo {author} {\bibfnamefont {X.-Q.}\ \bibnamefont {Luo}},
  \bibinfo {author} {\bibfnamefont {Y.-P.}\ \bibnamefont {Wang}}, \bibinfo
  {author} {\bibfnamefont {T.-F.}\ \bibnamefont {Li}}, \ and\ \bibinfo {author}
  {\bibfnamefont {J.}~\bibnamefont {You}},\ }\href@noop {} {\bibfield
  {journal} {\bibinfo  {journal} {Nature communications}\ }\textbf {\bibinfo
  {volume} {8}},\ \bibinfo {pages} {1368} (\bibinfo {year}
  {2017}{\natexlab{b}})}\BibitemShut {NoStop}%
\bibitem [{\citenamefont {Scully}\ and\ \citenamefont
  {Zubairy}(1997)}]{Scully:1601132}%
  \BibitemOpen
  \bibfield  {author} {\bibinfo {author} {\bibfnamefont {M.~O.}\ \bibnamefont
  {Scully}}\ and\ \bibinfo {author} {\bibfnamefont {M.~S.}\ \bibnamefont
  {Zubairy}},\ }\href {https://cds.cern.ch/record/1601132} {\emph {\bibinfo
  {title} {{Quantum optics}}}}\ (\bibinfo  {publisher} {Cambridge University
  Press},\ \bibinfo {address} {Cambridge},\ \bibinfo {year} {1997})\BibitemShut
  {NoStop}%
\bibitem [{\citenamefont {Lemonde}\ \emph {et~al.}(2014)\citenamefont
  {Lemonde}, \citenamefont {Didier},\ and\ \citenamefont
  {Clerk}}]{PhysRevA.90.063824}%
  \BibitemOpen
  \bibfield  {author} {\bibinfo {author} {\bibfnamefont {M.-A.}\ \bibnamefont
  {Lemonde}}, \bibinfo {author} {\bibfnamefont {N.}~\bibnamefont {Didier}}, \
  and\ \bibinfo {author} {\bibfnamefont {A.~A.}\ \bibnamefont {Clerk}},\ }\href
  {\doibase 10.1103/PhysRevA.90.063824} {\bibfield  {journal} {\bibinfo
  {journal} {Phys. Rev. A}\ }\textbf {\bibinfo {volume} {90}},\ \bibinfo
  {pages} {063824} (\bibinfo {year} {2014})}\BibitemShut {NoStop}%
\bibitem [{\citenamefont {Wang}\ \emph {et~al.}(2016)\citenamefont {Wang},
  \citenamefont {Miranowicz}, \citenamefont {Li},\ and\ \citenamefont
  {Nori}}]{PhysRevA.93.063861}%
  \BibitemOpen
  \bibfield  {author} {\bibinfo {author} {\bibfnamefont {X.}~\bibnamefont
  {Wang}}, \bibinfo {author} {\bibfnamefont {A.}~\bibnamefont {Miranowicz}},
  \bibinfo {author} {\bibfnamefont {H.-R.}\ \bibnamefont {Li}}, \ and\ \bibinfo
  {author} {\bibfnamefont {F.}~\bibnamefont {Nori}},\ }\href {\doibase
  10.1103/PhysRevA.93.063861} {\bibfield  {journal} {\bibinfo  {journal} {Phys.
  Rev. A}\ }\textbf {\bibinfo {volume} {93}},\ \bibinfo {pages} {063861}
  (\bibinfo {year} {2016})}\BibitemShut {NoStop}%
\bibitem [{\citenamefont {Xu}\ \emph {et~al.}(2019)\citenamefont {Xu},
  \citenamefont {Shi}, \citenamefont {Liao},\ and\ \citenamefont
  {Chen}}]{PhysRevA.100.053802}%
  \BibitemOpen
  \bibfield  {author} {\bibinfo {author} {\bibfnamefont {X.-W.}\ \bibnamefont
  {Xu}}, \bibinfo {author} {\bibfnamefont {H.-Q.}\ \bibnamefont {Shi}},
  \bibinfo {author} {\bibfnamefont {J.-Q.}\ \bibnamefont {Liao}}, \ and\
  \bibinfo {author} {\bibfnamefont {A.-X.}\ \bibnamefont {Chen}},\ }\href
  {\doibase 10.1103/PhysRevA.100.053802} {\bibfield  {journal} {\bibinfo
  {journal} {Phys. Rev. A}\ }\textbf {\bibinfo {volume} {100}},\ \bibinfo
  {pages} {053802} (\bibinfo {year} {2019})}\BibitemShut {NoStop}%
\bibitem [{\citenamefont {Xu}\ \emph {et~al.}(2016)\citenamefont {Xu},
  \citenamefont {Chen},\ and\ \citenamefont {Liu}}]{PhysRevA.94.063853}%
  \BibitemOpen
  \bibfield  {author} {\bibinfo {author} {\bibfnamefont {X.-W.}\ \bibnamefont
  {Xu}}, \bibinfo {author} {\bibfnamefont {A.-X.}\ \bibnamefont {Chen}}, \ and\
  \bibinfo {author} {\bibfnamefont {Y.-x.}\ \bibnamefont {Liu}},\ }\href
  {\doibase 10.1103/PhysRevA.94.063853} {\bibfield  {journal} {\bibinfo
  {journal} {Phys. Rev. A}\ }\textbf {\bibinfo {volume} {94}},\ \bibinfo
  {pages} {063853} (\bibinfo {year} {2016})}\BibitemShut {NoStop}%
\bibitem [{\citenamefont {Safavi-Naeini}\ \emph {et~al.}(2011)\citenamefont
  {Safavi-Naeini}, \citenamefont {Alegre}, \citenamefont {Chan}, \citenamefont
  {Eichenfield}, \citenamefont {Winger}, \citenamefont {Lin}, \citenamefont
  {Hill}, \citenamefont {Chang},\ and\ \citenamefont {Painter}}]{Nature2011}%
  \BibitemOpen
  \bibfield  {author} {\bibinfo {author} {\bibfnamefont {A.~H.}\ \bibnamefont
  {Safavi-Naeini}}, \bibinfo {author} {\bibfnamefont {T.~P.~M.}\ \bibnamefont
  {Alegre}}, \bibinfo {author} {\bibfnamefont {J.}~\bibnamefont {Chan}},
  \bibinfo {author} {\bibfnamefont {M.}~\bibnamefont {Eichenfield}}, \bibinfo
  {author} {\bibfnamefont {M.}~\bibnamefont {Winger}}, \bibinfo {author}
  {\bibfnamefont {Q.}~\bibnamefont {Lin}}, \bibinfo {author} {\bibfnamefont
  {J.~T.}\ \bibnamefont {Hill}}, \bibinfo {author} {\bibfnamefont {D.~E.}\
  \bibnamefont {Chang}}, \ and\ \bibinfo {author} {\bibfnamefont
  {O.}~\bibnamefont {Painter}},\ }\href@noop {} {\bibfield  {journal} {\bibinfo
   {journal} {Nature}\ }\textbf {\bibinfo {volume} {472}},\ \bibinfo {pages}
  {69} (\bibinfo {year} {2011})}\BibitemShut {NoStop}%
\bibitem [{\citenamefont {Clark}\ \emph {et~al.}(2017)\citenamefont {Clark},
  \citenamefont {Lecocq}, \citenamefont {Simmonds}, \citenamefont {Aumentado},\
  and\ \citenamefont {Teufel}}]{Clark}%
  \BibitemOpen
  \bibfield  {author} {\bibinfo {author} {\bibfnamefont {J.~B.}\ \bibnamefont
  {Clark}}, \bibinfo {author} {\bibfnamefont {F.}~\bibnamefont {Lecocq}},
  \bibinfo {author} {\bibfnamefont {R.~W.}\ \bibnamefont {Simmonds}}, \bibinfo
  {author} {\bibfnamefont {J.}~\bibnamefont {Aumentado}}, \ and\ \bibinfo
  {author} {\bibfnamefont {J.~D.}\ \bibnamefont {Teufel}},\ }\href@noop {}
  {\bibfield  {journal} {\bibinfo  {journal} {Nature}\ }\textbf {\bibinfo
  {volume} {541}},\ \bibinfo {pages} {191} (\bibinfo {year}
  {2017})}\BibitemShut {NoStop}%
\end{thebibliography}%

\end{document}